\documentclass[final,5p,times,twocolumn,authoryear]{elsarticle}

\journal{Astronomy $\&$ Computing}

\usepackage[T1]{fontenc}
\usepackage{hyperref}
\usepackage{graphicx}
\usepackage{amsmath}
\usepackage{amssymb}
\usepackage{newtxtext}
\usepackage[varvw]{newtxmath}
\usepackage{xcolor}
\usepackage{bm}
\usepackage[ruled,vlined]{algorithm2e}
\usepackage{multirow}
\usepackage{booktabs}
\usepackage{subcaption}
\usepackage{spverbatim}
\usepackage{listings}

\definecolor{yamlGreen}{RGB}{63,127,63}
\definecolor{yamlBlue}{RGB}{0,0,255}
\definecolor{yamlPurple}{RGB}{127,0,127}
\definecolor{yamlGray}{RGB}{100,100,100}

\lstdefinestyle{yaml}{
    language={},
    basicstyle=\ttfamily\scriptsize,
    numbers=left,
    numberstyle=\tiny\color{gray},
    stepnumber=1,
    numbersep=5pt,
    backgroundcolor=\color{gray!10},
    showspaces=false,
    showstringspaces=false,
    showtabs=false,
    frame=single,
    tabsize=2,
    captionpos=b,
    breaklines=true,
    breakatwhitespace=true,
    breakautoindent=true,
    linewidth=\textwidth,
    moredelim=[l][\color{yamlGray}]{\#},
    moredelim=[l][\color{yamlPurple}]{---},
    moredelim=**[s][\color{yamlBlue}]{*}{*},
    morestring=[b]',
    morestring=[b]"
}

\newcommand{\yamlfromfile}[1]{
    \lstinputlisting[style=yaml]{#1}
}

\definecolor{commentColor}{RGB}{0,128,0}
\definecolor{stringColor}{RGB}{163,21,21}
\definecolor{backgroundGray}{RGB}{245,245,245}

\lstdefinestyle{bash}{
    language=bash,
    basicstyle=\ttfamily\small,
    numbers=none,
    commentstyle=\color{commentColor},
    stringstyle=\color{stringColor},
    showspaces=false,
    showstringspaces=false,
    frame=none,
    backgroundcolor=\color{backgroundGray},
    breaklines=true,
    postbreak=\mbox{\textcolor{gray}{$\hookrightarrow$}\space}
}

\newcommand{\prox}{\mbox{prox}}
\newcommand{\M}{\bm{M}}
\newcommand{\B}{\bm{B}}
\newcommand{\V}{\bm{V}}
\newcommand{\J}{\bm{J}}
\newcommand{\T}{\bm{T}}
\newcommand{\I}{\bm{I}}
\newcommand{\R}{\bm{R}}
\newcommand{\C}{\bm{C}}
\newcommand{\X}{\bm{X}}

\newcommand{\qcal}{\textsc{QuartiCal}}
\newcommand{\stimela}{\textsc{Stimela2}}
\newcommand{\pfb}{\textsc{pfb-imaging}}

\newcommand{\dask}{\textsc{Dask}}
\newcommand{\zarr}{\textsc{Zarr}}

\newcommand{\daskms}{\textsc{Dask-MS}}
\newcommand{\codexafricanus}{\textsc{codex africanus}}
\newcommand{\numpy}{\textsc{NumPy}}
\newcommand{\numba}{\textsc{Numba}}

\newcommand{\scipy}{\textsc{SciPy}}

\newcommand{\xarray}{\textsc{XArray}}
\newcommand{\wsclean}{\textsc{WSClean}}

\newcommand{\ducc}{\textsc{ducc}}
\newcommand{\python}{\textsc{Python}}

\newcommand{\africanus}{\textsc{Africanus}}
\newcommand{\xova}{\textsc{xova}}
\newcommand{\click}{\textsc{Click}}
\newcommand{\kubernetes}{\textsc{Kubernetes}}

\newcommand{\init}{\texttt{init}}
\newcommand{\grid}{\texttt{grid}}
\newcommand{\modeltocomps}{\texttt{model2comps}}
\newcommand{\degrid}{\texttt{degrid}}
\newcommand{\restore}{\texttt{restore}}
\newcommand{\sara}{\texttt{sara}}
\newcommand{\fluxmop}{\texttt{fluxtractor}}

\newcommand{\wgridder}{\texttt{wgridder}}
\newcommand{\clean}{CLEAN}

\newcommand{\aws}{AWS}
\newcommand{\eso}{ESO137}

\begin{document}
\begin{frontmatter}

\title{Africanus III. \pfb\ - a flexible radio interferometric imaging suite}

\begin{abstract}
The popularity of the CLEAN algorithm in radio interferometric imaging stems from its maturity, speed, and robustness. While many alternatives have been proposed in the literature, none have achieved mainstream adoption by astronomers working with data from interferometric arrays operating in the big data regime. This lack of adoption is largely due to increased computational complexity, absence of mature implementations, and the need for astronomers to tune obscure algorithmic parameters.

This work introduces \pfb: a flexible library that implements the scaffolding required to develop and accelerate general radio interferometric imaging algorithms. We demonstrate how the framework can be used to implement a sparsity-based image reconstruction technique known as (unconstrained) SARA in a way that scales with image size rather than data volume and features interpretable algorithmic parameters. The implementation is validated on terabyte-sized data from the MeerKAT telescope, using both a single compute node and Amazon Web Services computing instances.
\end{abstract}

\author[sarao,ratt]{H.~L.~Bester}\ead{lbester@sarao.ac.za}
\author[ratt]{J.~S.~Kenyon}
\author[math,maxwell,basp]{A.~Repetti}
\author[sarao]{S.~J.~Perkins}
\author[ratt,sarao,ira]{O.~M.~Smirnov}
\author[sarao]{T.~Blecher}
\author[listic]{Y.~Mhiri}
\author[mpa,tig]{J.~Roth}
\author[oxford,btl,ratt,sarao]{I.~Heywood}
\author[basp]{Y.~Wiaux}
\author[sarao]{B.~V.~Hugo}

\affiliation[sarao]{organization={South African Radio Astronomy Observatory (SARAO)},
            city={Cape Town},
            postcode={7700},
            state={Western Cape},
            country={South Africa}}

\affiliation[ratt]{organization={Centre for Radio Astronomy Techniques \& Technologies (RATT),
Department of Physics and Electronics, Rhodes University},
            city={Makhanda},
            postcode={6139},
            state={Eatern Cape},
            country={South Africa}}

\affiliation[ira]{organization={Institute for Radioastronomy, National Institute of Astrophysics (INAF IRA)},
            city={Bologna},
            postcode={40129},
            country={Italy}}

\affiliation[maxwell]{organization={Maxwell Institute for Mathematical Sciences},
            city={Edinburgh},
            postcode={EH9 3FD},
            country={United Kingdom}}

\affiliation[math]{organization={School of Mathematical and Computer Sciences, Heriot-Watt University},
            city={Edinburgh},
            postcode={EH14 4AS},
            country={United Kingdom}}

\affiliation[basp]{organization={Institute of Sensors, Signals and Systems, Heriot-Watt University},
            city={Edinburgh},
            postcode={EH14 4AS},
            country={United Kingdom}}

\affiliation[mpa]{organization={Max Planck Institute for Astrophysics}, 
            city={Garching},
            postcode={85748},
            country={Germany}}

\affiliation[tig]{organization={Technische Universität M\"unchen (TUM)},
            city={Garching},
            postcode={85748},
            country={Germany}}

\affiliation[listic]{organization={Laboratoire d'Informatique, Systèmes, Traitement de l'Information et de la Connaissance (LISTIC), Université Savoie Mont Blanc (USMB)},
            city={Annecy},
            postcode={74940},
            country={France}}

\affiliation[oxford]{organization={Astrophysics, Department of Physics, University of Oxford},
            city={Oxford},
            postcode={OX13RH},
            country={United Kingdom}}

\affiliation[btl]{organization={Breakthrough Listen, Astrophysics, Department of Physics, University of Oxford},
            city={Oxford},
            postcode={OX13RH},
            country={United Kingdom}}

\begin{keyword}
    standards -- techniques: interferometric --
    standards -- techniques: image processing --
    Computer systems organization: Pipeline computing --
    Software and its engineering: Data flow architectures --
    Software and its engineering: Cloud computing --
    Software and its engineering: Interoperability
\end{keyword}

\end{frontmatter}

\section{Introduction}
\label{sec:intro}
Radio astronomy is in an era of massive expansion. With modern telescopes such as LOFAR \cite{vanhaarlem2013}, MeerKAT \cite{jonas2016meerkat} and ASKAP \cite{Hotan_2021} already in operation, and with the upcoming Square Kilometre array (SKA) \cite{schilizzi2008square}, next generation Very Large Array (ngVLA) \cite{ngvla} and the Deep Synoptic Array (DSA) \cite{dsa2000}, it is more important than ever to scrutinize the techniques and technologies required to reliably process and store the vasts amounts of data expected over the next few decades. This is especially true for projects aiming to extract science close to the noise floor of the observation.

Astrophysical signals experience a number of physical transformations as they travel to detectors placed on earth. Radio interferometers measure the coherency of the signal's electric field by correlating the voltages it induces across a pair of antenna receivers. The radio interferometry measurement equation (RIME) (see \cite{hamaker1996understanding1, hamaker1996understanding2, smirnov2011revisiting1, smirnov2011revisiting2} for example) is a mathematical model of the transformations an astrophysical signal undergoes as it propagates from its source, through antenna receivers, to the point that the signals from two antennas are correlated to form the data products (viz. visibilities) that are then stored for further processing. Each physical transformation can be encoded mathematically as a complex $2\times2$ Jones matrix \cite{smirnov2011revisiting1} with exact parametrisation dependent on the nature of the transformation. Multiple transformations of the signal can be encoded as a chain of such matrices acting in the correct order. One of the key challenges in processing interferometric data is producing physically realistic images in the presence of such Jones matrices (also referred to as calibration parameters or gains). Since instrumental and atmospheric effects (encoded as Jones matrices) are not usually not known in advance, they have to be inferred alongside the signal of interest. In addition, since a finite collection of antennas can only synthesize a partially filled aperture, the process of reconstructing an image from raw visibilities is ill-posed i.e. it is not unique. Ill-posed inverse problems are not uncommon in science and engineering applications and there exists a plethora of literature on the subject (for example, see \cite{illposed2019} and references therein). In particular, Bayes' law gives the probability distribution governing the signal (i.e. the posterior) under a particular measurement model, given our prior assumptions. Importantly, since the problem is ill-posed, the posterior is completely degenerate without suitable prior assumptions.

Unfortunately, because of the complexity and scale of the problem, a full Bayesian treatment remains elusive (see however \cite{resolve1,sutter2014,Cai2018I,Arras_2021,roth2024fastresolvefastbayesianradio,liaudat2024}) and we have to content ourselves with point (or maximum a posteriori (MAP)) estimates obtained by maximising the joint probability of data and signal. Such approaches can leverage optimisation theory to efficiently deal with the typically high dimensional inference problems encountered in radio interferometry. Even then, simultaneously inferring both calibration parameters and signal is not usually computationally feasible. For this reason, the solution to the ill-posed inverse problem has historically been sought by separating the problem into separate calibration (i.e. estimation of the Jones matrices) and imaging (i.e. estimation of the sky brightness distribution) problems. The framework considered in this paper is no exception.

This separation simplifies the problem for a number of reasons. Of particular relevance to the current paper series is the fact that the two problems parallelise completely differently and have very different memory consumption patterns, making it much simpler to efficiently distribute the two problems separately. Although much of the discussion that follows is relevant to both the imaging and calibration problems, this work is mainly concerned with estimating the sky brightness distribution in the presence of known (direction independent) Jones matrices (further details regarding the calibration problem are given in the accompanying paper \cite{africanus2}). We propose an approach based on the preconditioned forward-backward (PFB) algorithm developed in \cite{pfb3} and detail an implementation based on the technologies and ideologies discussed in the first paper in this series \cite{africanus1}. Our proposal is motivated by the fact that the most ubiquitous imaging algorithm in radio interferometry, viz. \clean\ \citep[see e.g.][]{clean1974,clean2011,Offringa_2014}, suffers from a number of limitations.

Possibly the biggest limitation of \clean\ is the fact that it is not formulated as an optimisation algorithm rooted in forward modelling but rather as a procedure aimed at minimising discrepancies between the data and a model composed of fairly simplistic (e.g. Dirac delta and Gaussian) components. This is suboptimal for a number of reasons. Firstly, the procedure does not produce physically realistic reconstructions of the sky, especially for diffuse emission. This is one of the primary reasons why astronomers don't typically use model images produced by \clean\ to infer morphological properties of celestial sources, preferring to use restored images (see \ref{sec:restore}) instead. This gives rise to the notion that resolution has to be traded against sensitivity. In what follows we elaborate on why this is a flawed notion and show that algorithms can be designed to deliver comparable, or even superior, resolution than the more uniformly weighted images produced by \clean.

Secondly, ill-posed inverse problems require regularisation for a unique inversion. \clean\ encodes this regularisation into the structure of the algorithm in a way that makes it very difficult to ascertain exactly what objective function is being minimised. This makes it very difficult to reason about uncertainties in the reconstructed images in a statistically robust way. While this work does not deliver a satisfactory mechanism for quantifying uncertainties, formulating the inverse problem as a statistical optimisation problem is a necessary step in that direction. Such an approach lends itself more readily to algorithms that enable some form of uncertainty quantification and makes it easier to account for systematics like the primary beam in a more robust way than is possible with \clean.

For all its faults, \clean\ is remarkably fast and stands out for its ease of parameter tuning, both features which have contributed to its widespread adoption in radio interferometric imaging. One reason for this is that the algorithm's parameters become intuitively understandable once you are familiar with how \clean\ operates. This is facilitated by the manner in which parameters, such as the stopping criteria, are defined based on characteristics of the residual image. The algorithmic parameters that need to be specified to successfully deploy alternatives to \clean\ are often more obscure and can be hard for new users to come to grips with. This, in combination with increased computational complexity, has likely contributed to their relatively slow uptake by the community, despite their transformative potential.

Our goals for the paper are threefold. First and foremost, we develop a flexible and generalisable framework rooted in forward modelling that is suitable for implementing and accelerating imaging algorithms posed as statistical optimisation problems. One of the principal motivations for creating this framework, dubbed \pfb{}, is to alleviate the burden on developers aiming to test new imaging algorithms on large interferometric data sets. As discussed in $\S$~\ref{sec:software}, the \pfb\ framework achieves this by deconstructing the main steps in a typical imaging workflow into separate applications which interact with simpler and more efficient data structures. Secondly, we provide an example of using the framework to implement and accelerate a variant of the Sparsity Averaging Re-weighted Analysis (SARA) family of algorithms \citep[see e.g.][]{sara,hypersara, thouvenin2023,wilber2023}. We show how the hyper-parameters of the algorithm can be coaxed into an intuitive form and provide some guidelines for setting them when dealing with real data. Lastly, we show how our imaging framework ties in with other technologies in the \africanus\ ecosystem \cite{africanus1,africanus2,africanus4} and evaluate some of its scaling characteristics in the distributed setting.

The paper is organised as follows. In $\S$~\ref{sec:measmod} we review the form of the measurement operator in the presence of known direction independent Jones matrices and show how data can be cast into a convenient form for imaging Stokes parameters. We also discuss the main insight underpinning the \clean\ algorithm and use it to develop some intuition for the imaging problem. In $\S$~\ref{sec:inverse} we outline the mathematical details behind the proposed optimisation framework and discuss its implications for processing radio interferometric data. We then develop some preconditioning strategies that can be used to accelerate the convergence of general imaging algorithms and apply the framework to a variant of the SARA family of algorithms. $\S$~\ref{sec:software} presents the \pfb\ software package responsible for the numerical implementation of the material presented in this paper. Many of the practical considerations involved when dealing with large interferometric data sets are discussed in this section. $\S$\ref{sec:results} demonstrates an application of \pfb\ to MeerKAT data. Our implementation is benchmarked against the popular imaging library \wsclean\ \cite{Offringa_2014,offringa2017multiscale} on a single node. We also illustrate some of its scaling characteristics in the distributed setting by running it on Amazon Web Services (\aws) instances. Finally, we conclude with a discussion and some future prospects in $\S$~\ref{sec:conclusion}.

\section{Measurement model}
\label{sec:measmod}
In this section, we discuss the interferometric measurement model. We start by considering the problem generally in the presence of all polarisation products and show how the data can be cast into a format that is convenient for imaging. This is compared to the conventional approach of forming corrected data by applying the inverse of the gains to data. We further detail the transformations required to write the inference problem in terms of Stokes parameters and discuss some of the advantages and disadvantages of transforming the data into this form. We then specialise to total intensity (i.e. Stokes $I$) imaging and derive the operators required to perform this typically high dimensional optimisation problem efficiently.

\subsection{Stokes visibilities}
Neglecting direction dependent effects (DDEs), the apparent sky brightness distribution, $\B = \B(l, m, \nu)$, can be related to the visibilities, $\V_{pq} = \V_{pq}(t, \nu)$ according to
\begin{equation}
    \V_{pq} = \M_{pq} \int \B e^{-2\pi\imath \frac{\nu}{c} (u_{pq}l + v_{pq}m + w_{pq}(n-1))} \frac{dl dm}{n},
    \label{RIME}
\end{equation}
where $\left(l,m,n = \sqrt{1 - l^2 - m^2}\right)$ are direction cosines on the unit sphere with $(l=0, m=0)$ aligned to the tracking center of the interferometer, $(u_{pq},v_{pq},w_{pq})$ give the relative positions between antennas $p$ and $q$ in a frame such that $u_{pq}$ and $v_{pq}$ span the plane orthogonal to $n$, $\nu$ denotes the frequency of the observation, $c$ is the speed of light and $t$ denotes proper time in the frame of the telescope. Note that the baseline coordinates, $(u_{pq},v_{pq},w_{pq})$, are time dependent due to the peculiar motion of the earth. The quantity $M_{pq} = M_{pq}(t, \nu)$ is a Mueller-like term \citep[see e.g.][]{Goldstein2003PolarizedLight} representing the combined effect of all Jones matrices relevant to antenna $p$ and $q$. Once the Jones chains for each antenna have been combined into single effective Jones term, $\J_p$ for antenna $p$ say, $\M_{pq}$ can be written as $\M_{pq} = \J_q^* \otimes \J_p$ where $\otimes$ denotes the Kronecker product and a superscript ${}^*$ denotes complex conjugation. The RIME can be written in the more compact form $\V_{pq} = \M_{pq} \X_{pq}$ by denoting the coherencies associated with elements of the sky brightness distribution as
\begin{equation}
    \X_{pq} = \int \B e^{-2\pi\imath \frac{\nu}{c} (u_{pq}l + v_{pq}m + w_{pq}(n-1))} \frac{dl dm}{n}.
    \label{ccoh}
\end{equation}
This treats $\X_{pq}$ as a complex $4 \times 1$ vector with entries for each of the correlation products and $\M_{pq}$ as a $4 \times 4$ Mueller matrix, both as continuous functions of $(t, \nu)$. At any  $(l,m,\nu)$, the $4 \times 1$ vector $\B$ contains correlation products which can be related to the Stokes parameters, $\I = [I, Q, U, V]^T$, via the constant matrix $\T$. The exact form of $\T$ depends on the feed type, i.e. linear or circular, used in the antenna receivers. In the case of linear feeds, and with $\I$ defined as above, we have
\begin{equation}
    \B = \T \I, \quad \mbox{where} \quad
    \B = \begin{bmatrix}
    I + Q \\
     U - \imath V \\
     U + \imath V \\
     I - Q
    \end{bmatrix}, \quad
    \T = \begin{bmatrix}
    1 & 1 & 0 & 0 \\
    0 & 0 & 1 & -\imath \\
    0 & 0 & 1 & \imath \\
    1 & -1 & 0 & 0
    \end{bmatrix}.
\end{equation}
Denoting the integral operator (sometimes referred to as the degridding operator) associated with antenna $p$ and $q$ as $\R_{pq}$, it is possible to write \eqref{RIME} in the form
\begin{equation}
    \V_{pq} = \M_{pq} \R_{pq} \T \I + \bm{\epsilon}_{pq}, \quad \mbox{where} \quad \bm{\epsilon}_{pq} \sim \mathcal{N}(0, \bm{\Sigma}_{pq}),
    \label{measmodpq}
\end{equation}
where we have indulged in a slight abuse of notation by redefining $\V_{pq}$ as the noisy visibilities. In what follows we assume that $\bm{\epsilon}_{pq}$ is a realisation of proper (or circular) complex noise with diagonal covariance matrix $\bm{\Sigma}_{pq}$. Together with the measurement operator, this gives the negative log-likelihood (i.e. the data fidelity term) as
\begin{equation}
    f(\I) = (\V_{pq} - \M_{pq} \R_{pq} \T \I)^\dagger \bm{\Sigma}_{pq}^{-1} (\V_{pq} - \M_{pq} \R_{pq} \T \I).
    \label{muelchi2}
\end{equation}
This formulation allows us to formulate an inference problem in terms of the Stokes parameters which is a common way to encode physical information about astrophysical sources.

It is possible to go one step further and write \eqref{muelchi2} directly in terms of the Stokes coherencies defined as $\C_{pq} = \R_{pq} \I$\footnote{We use the term Stokes coherencies to distinguish these quantities from the coherencies associated with elements of $\B$ as in \eqref{ccoh}. The discussion that follows can equivalently be formulated in terms of $\X_{pq}$ but it is more convenient for our purposes to formulate it in terms of $\C_{pq}$.}. Since the integral operator in \eqref{RIME} is the same for each correlation, $\R_{pq}$ commutes with $\T$ and we can write \eqref{muelchi2} as
\begin{equation}
    f(\C_{pq}) = (\V_{pq} - \M_{pq} \T \C_{pq})^\dagger \bm{\Sigma}_{pq}^{-1} (\V_{pq} - \M_{pq} \T \C_{pq}).
\end{equation}
This is a quadratic in $\C_{pq}$ so it is possible to write it equivalently as
\begin{equation}
    f(\C_{pq}) \overset{\triangle}{=}(\hat{\C}_{pq} - \C_{pq})^\dagger \mathcal{W}_{pq} (\hat{\C}_{pq} - \C_{pq}),
\end{equation}
with suitable definitions of $\hat{\C}_{pq}$ and $\mathcal{W}_{pq}$. The notation $\overset{\triangle}{=}$ introduced above denotes equality up to an additive constant that is independent of the argument of the function. By quadratic completion, it can be shown that
\begin{equation}
    \hat{\C}_{pq} = \left(\T^\dagger \M_{pq}^\dagger \bm{\Sigma}_{pq}^{-1} \M_{pq} \T\right)^{-1} \T^\dagger \M_{pq}^\dagger \bm{\Sigma}_{pq}^{-1} \V_{pq},
    \label{cdata}
\end{equation}
and
\begin{equation}
    \mathcal{W}_{pq} = \T^\dagger \M_{pq}^\dagger \bm{\Sigma}_{pq}^{-1} \M_{pq} \T,
    \label{cweight}
\end{equation}
are the desired definitions. In what follows we refer to $\hat{\C}_{pq}$ and $\mathcal{W}_{pq}$ as the observed Stokes coherencies and Mueller weights respectively. Under the assumption that $\bm{\epsilon}_{pq} \sim \mathcal{N}(0, \bm{\Sigma}_{pq})$, these expressions provide a statistically consistent way to infer Stokes parameters directly from $\hat{\C}_{pq}$ instead of $\V_{pq}$. In other words, by using \eqref{cdata} and \eqref{cweight} to precompute $\hat{\C}_{pq}$ and $\mathcal{W}_{pq}$ respectively, it is possible to form the new data fidelity term
\begin{equation}
    f(\I) = (\hat{\C}_{pq} - \R_{pq} \I)^\dagger \mathcal{W}_{pq} (\hat{\C}_{pq} - \R_{pq} \I),
    \label{cchi2}
\end{equation}
and perform inference on $\I$ directly without the need to reapply the gains at each iteration. This is convenient when attempting to reconstruct $\I$ in the presence of known Jones matrices. Note, however, that $\mathcal{W}_{pq}$ is a $4 \times 4$ matrix with entries strongly dependent on the magnitude of the off-diagonal elements of $\J_p$ and $\J_q$. Since telescopes are often engineered to keep these off-diagonal (leakage) terms small, $\mathcal{W}_{pq}$ can often be well approximated by a diagonal matrix. By discarding the off diagonal elements of $\mathcal{W}_{pq}$ we are essentially incorrectly assuming that the Stokes parameters are statistically independent. This might have detrimental consequences for full polarisation imaging which is, unfortunately, beyond the scope of the current work. We therefore use the diagonal entries of $\mathcal{W}_{pq}$ to obtain the weights corresponding to the Stokes coherencies. These are also used to construct point spread functions (PSFs) for the individual Stokes parameters. As discussed in $\S$~\ref{sec:precond}, this is useful for developing preconditioning strategies to accelerate the convergence rate of certain classes of algorithms.

Finally, once the data have been transformed into Stokes form \eqref{cchi2}, it becomes far more amenable to averaging. This holds because $\hat{\C}_{pq}$ has been corrected for the time and frequency response of the instrument, which varies more rapidly than the astrophysical sources of interest. The fact that the intermediary data products $\hat{\C}_{pq}$ and $\mathcal{W}_{pq}$ required for imaging can do away with much of the metadata required by other applications in the pipeline (e.g. antenna labels are needed for subsequent self-calibration but not for simple imaging) also makes it possible to use baseline dependent averaging (BDA) effectively. The intermediary data products can therefore be orders of magnitude smaller than the original data, depending on the degree of averaging and the required number of Stokes products. This should be contrasted to the more usual approach of applying the inverse of the gains to the data to derive the so called corrected data and weights, which effectively doubles the size of the data that needs to be stored.

\subsection{Imaging}
\label{sec:imaging}
The imaging problem simplifies quite drastically if we are only interested in a single Stokes data product. For simplicity, we restrict the current discussion to total intensity imaging. In this case, we need only consider the first elements\footnote{Similar expressions follow for the other Stokes parameters by considering only the diagonal entries of \eqref{cweight}.} of \eqref{cdata} and \eqref{cweight}. These give the scalar Stokes $I$ data and weights for each baseline at a particular time and frequency. Information from multiple baselines can be combined by stacking visibilities into a vector, $y$, and by combining the measurement operator for multiple baselines into the (rank deficient) matrix $R$. Further discretising the total intensity sky brightness distribution into a pixelated image, $x$, gives a measurement model of the form
\begin{equation}
    y = R x + \epsilon, \quad \epsilon \sim \mathcal{N}\left(0, \Sigma \right),
    \label{discRIME}
\end{equation}
where $\Sigma$ is the diagonal covariance matrix formed by stacking the inverse weights extracted from the first elements of \eqref{cweight} onto the diagonal. Implicit in the above is the assumption that a single image maps to all frequencies in the output. Since the image also varies with frequency, there is actually a measurement model such as \eqref{discRIME} for each imaging band so that $x \in \mathbb{R}^{n_b \times n_p}$ with $n_b$ the number of imaging bands and $n_p$ the number of spatial pixels.

For continuum imaging, it is typical to reconstruct the image at a much lower frequency resolution than that of the data. The measurement operator is therefore a linear mapping
\begin{equation}
    R \colon \mathbb{R}^{n_b \times n_p} \to \mathbb{C}^{n_\nu \times n_r}
    \label{immeasop}
\end{equation}
where $n_\nu \gg n_b$ is the number of frequency channels and $n_r$ the number of Fourier measurements per channel (i.e. the number of time integrations multiplied by the number of baselines). Since the image is defined on a regular grid, and is typically much smaller than the data, it is advantageous to formulate the problem in image space as far as possible. As shown below, the likelihood for the problem can be approximately cast into image space by exploiting the fact that the mapping \eqref{immeasop} reduces to a non-uniform two dimensional Fourier transform for each imaging band when the array is coplanar, or when we are only interested in a small patch of the sky \citep[see e.g.][]{cornwell1992radio}. In what follows we refer to this regime as the coplanar array limit. Note that, since all linear operators considered in this work operate on bands independently, we simplify the notation by leaving the dependence on imaging band implicit throughout. The only exception is made for the function defined by \eqref{gfunc} which necessitates communication across imaging bands.

The coarse discretisation of the measurement operator \eqref{immeasop} along the frequency axis can be detrimental since it essentially results in step-wise constant models. Some imaging applications (see e.g. the \textsf{deconvolution-channels} parameter in \wsclean\ \cite{Offringa_2014, offringa2017multiscale}) allow refining the frequency resolution of the model by interpolating in frequency (by fitting a polynomial to the non-zero components, for example) and using a more accurate degridding operator. This can be rather confusing since optimisation algorithms typically require the measurement operator to be consistent for convergence. In other words, given a random sample from the domain of $R$, $\xi_1$ say, and one from its co-domain, $\xi_2$ say, the real and imaginary parts of the inner product $\xi_2^\dagger R \xi_1$, being scalar, should be invariant under a conjugate transpose operation i.e.
\begin{equation}
    \mbox{real}\left(\xi_2^\dagger R \xi_1\right) = \mbox{real}\left(\xi_1^\dagger R^\dagger \xi_2\right),
    \label{adjointness}
\end{equation}
and similarly for the imaginary part. \clean\ gets away with this inconsistency because of the implicit manner in which regularisation is encoded in the structure of the algorithm. It is much harder to do this in the presence of non-linearities or explicit constraints (e.g. positivity) because designing robust interpolators which respect such constraints is non-trivial. Since the imaging algorithm employed in this work requires the measurement operator to be consistent, we'll assume that it abides by \eqref{adjointness} throughout. Nevertheless, as discussed $\S$~\ref{sec:degrid}, up-sampling in frequency can be performed after deconvolution by using suitable interpolators.

The Gaussian nature of the noise gives the data fidelity term as
\begin{eqnarray}
    f(x) &=& (y - Rx)^\dagger \Sigma^{-1} (y - Rx), \\
     &\overset{\triangle}{=}& x^\dagger\left(R^\dagger \Sigma^{-1} R x - 2 I^D\right),
    \label{imagelik}
\end{eqnarray}
where the second equality follows from dropping terms independent of $x$ and $I^D = R^\dagger \Sigma^{-1} y$ denotes the dirty image. As mentioned above, the mapping \eqref{immeasop} approximates a two dimensional Fourier transform in the coplanar array limit. Thus, in this limit, the Fourier convolution theorem can be used to approximate the Hessian of \eqref{imagelik} as a convolution with the PSF of the instrument i.e.
\begin{equation}
    \nabla^2 f(x) = R^\dagger \Sigma^{-1} R x \approx I^{\mathrm{PSF}} * x
    = Z^\dagger F^\dagger \hat{I}^{\mathrm{PSF}} F Z x,
    \label{hessapprox}
\end{equation}
where $*$ denotes the 2D convolution, $Z$ is the zero padding operator, $F$ denotes the fast Fourier transform (FFT) and $\hat{I}^{\mathrm{PSF}}$ is a diagonal operator containing $F I^{\mathrm{PSF}}$ on its diagonal where $I^{\mathrm{PSF}}$ is the PSF computed on a grid the size of the padded image. Maximum accuracy is obtained when padding by a factor of two. This approximation, which is exact\footnote{Exact in the sense that it is accurate at least up to the accuracy of the gridding implementation if the image has been sufficiently over-sampled.} in the coplanar array limit, lies at the heart of the \clean\ algorithm \cite{clean1974, clean2011}. Under this approximation, assuming a noise free observation, we find that
\begin{equation}
    \nabla f(x) = 0 \quad \rightarrow \quad I^D = I^{\mathrm{PSF}} * x,
    \label{dconvm}
\end{equation}
i.e. the dirty image is the model convolved by the PSF. Note that, when the coplanar array limit does not apply, the imaging problem is not really a deconvolution problem because the PSF becomes direction dependent. Modern versions of the \clean\ algorithm address this issue by introducing major cycles i.e. revisiting the visibility data from time to time to evaluate the gradient exactly as
\begin{eqnarray}
    \nabla f(x) &=& 2 R^\dagger \Sigma^{-1}(Rx - y), \\
     &=& 2 R^\dagger \Sigma^{-1} R x - 2 I^D = -2 I^R,
    \label{visgrad}
\end{eqnarray}
where $I^R$ is called the residual image. $\S$~\ref{sec:inverse} formalises this idea and uses it to accelerate the convergence of more general imaging algorithms (see also \cite{bester2021}). It will be beneficial for the remainder of the paper to consider the approximation \eqref{hessapprox} in light of \eqref{visgrad} more carefully.

Putting aside non-coplanar array effects for the moment, consider that the operator $R$ and its adjoint can be implemented as a non-uniform FFT \citep[see e.g.][]{barnett_kernel} which uses the concept of convolutional gridding combined with the FFT for efficiency. While this initially seems like a purely technical detail, it forces one to think in terms of the Fourier (or $uv$) grid that measurements will be accumulated (i.e. weighted and convolved) onto. For the gridding operation (i.e. $R^\dagger$), each $uv$-cell will have a number of weighted data points accumulated onto it with that number depending on the sampling density in the neighbourhood of the $uv$-cell. Once this convolutional operation is done, the FFT (and technically also a grid corrector) is applied to map the Fourier grid into image space. This operation does not account for the fact that some $uv$-cells will accumulate many more samples than others, leading to a scale bias in image space. Since interferometric sampling patterns are typically core dominated, this bias manifests as large scales being over-emphasised in the gradient computed using \eqref{visgrad}.

\begin{figure*}
    \centering
    \includegraphics[width=2\columnwidth]{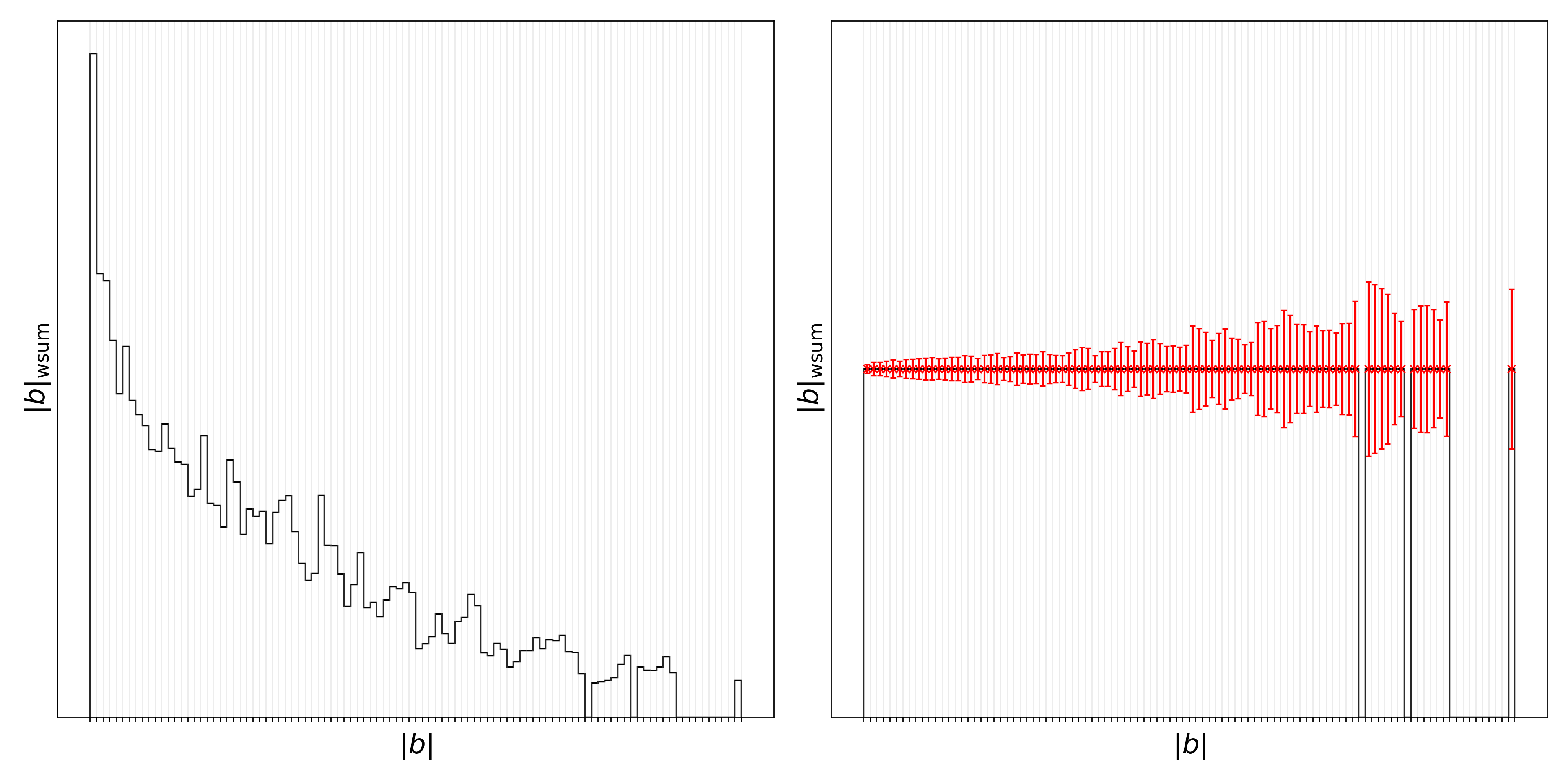}
    \caption{Sum of the weights $w_{\mathrm{sum}}$ as a function of baseline length $|b|$. The image on the left corresponds to the weights accumulated in each bin under natural weighting. It is difficult to assign error bars to each bin because the weights have been accumulated without normalising by the sum of the weights. The image on the right corresponds to uniform weighting. The error bars show the standard deviation after taking the weighted sum (including a normalisation by the sum of the weights). Standard deviations are computed as the inverse of the square root of the sum of the weights in the corresponding bins in the image on the left.}
    \label{fig:weighting}
\end{figure*}

It is common practice in radio interferometry to attempt to correct for this bias by introducing certain visibility weighting schemes \citep[see e.g.][]{briggs1995high}. The simplest and most intuitive being uniform weighting which can be understood by considering two passes through the data. In the first, one simply runs through all the data keeping track of the weights that are accumulated into each $uv$-cell, let us call this $uv_{\mathrm{wsum}}$. In the second stage, as the data are accumulated onto the grid, the weights corresponding to data points that fall within each cell are scaled by the inverse of $uv_{\mathrm{wsum}}$, resulting in a uniformly weighted grid. Not performing this normalisation is akin to computing a weighted sum without normalising by the sum of the weights. Uniform weighting attempts to give all scales equal emphasis but it does not propagate uncertainties on the data to uncertainties on the weighted grid (see also the discussion in the last paragraph of $\S$~\ref{sec:deconv}). The situation is depicted in Figure~\ref{fig:weighting}. Just because all scales in a uniformly weighted image have equal emphasis does not mean that they have equal uncertainties. The standard deviation on the weighted grid is inversely proportional to the square root of the weights used to normalise the weighted sum. Scales that have been sampled less are more uncertain.

Uniform weighting can be conceptualised in another way, one which will naturally lead us to the preconditioning strategies detailed in $\S$~\ref{sec:precond}. This is easiest to see if we assume periodic boundary conditions (so that the zero padding operators in \eqref{hessapprox} can be ignored) and invert \eqref{dconvm} as
\begin{equation}
    x = \left(F^\dagger \hat{I}^{\mathrm{PSF}} F\right)^{-1} I^D = F^\dagger (\hat{I}^{\mathrm{PSF}})^{-1} F I^D.
    \label{xpsfinvd}
\end{equation}
This expression has a pleasingly intuitive interpretation. Keeping in mind that the diagonal of $\hat{I}^{\mathrm{PSF}}$ essentially corresponds to the gridded weights, and $F I^D$ the gridded data, we see that applying the inverse of this convolution operator basically corrects for the sampling density of the interferometer. In practice, however, there are a number of reasons why the expression \eqref{xpsfinvd} is not very useful by itself. Firstly, the ill-posed nature of the problem means that some of the $uv$-cells may be empty making a direct inversion impossible. Secondly, assuming periodic boundary conditions can introduce unphysical artefacts, especially if there are bright sources close to the edge of the field of view. In addition, even if the array were coplanar, $\hat{I}^{\mathrm{PSF}}$ can contain both negative and imaginary values\footnote{For a coplanar array $\hat{I}^{\mathrm{PSF}}$ will only be real valued if the discrete Fourier transform is performed on an odd grid. For a non-coplanar array $\hat{I}^{\mathrm{PSF}}$ inevitably contains imaginary values if the field of view is large enough.} meaning that the linear operator approximating the Hessian in \eqref{hessapprox} is neither positive definite nor Hermitian. Lastly, given that our aim is to derive physically plausible images that are compatible with the data, physical constraints should also be enforced while solving the inverse problem.

Nevertheless, the expression \eqref{xpsfinvd} is not without merit. It lends a lot of intuition that can be utilised to better understand the problem at hand, at least in some qualitative way. For example, under the assumption of periodic boundary conditions, the discrete Fourier transform (DFT) provides an approximate eigenvalue decomposition of the Hessian operator in \eqref{hessapprox} with eigenvalues corresponding to the Fourier transform of the PSF. The largest eigenvalue is therefore proportional to the cell with the largest $uv_{\mathrm{wsum}}$. As we'll see in $\S$~\ref{sec:inverse}, this provides some insight into why more uniformly weighted problems have better convergence properties compared to their naturally weighted counterparts.

\section{Solving the inverse problem}
\label{sec:inverse}
This section details our general framework for deriving maximum a posteriori (MAP) solutions for the radio interferometric imaging problem. We'll start by outlining the specific PFB optimisation algorithm utilised in this work before moving on to the specific preconditioning and regularisation strategies used for inference.

\subsection{Optimisation framework}
\label{sec:pfb}
MAP estimation typically involves minimising objective functions of the form
\begin{equation}
    \Phi(x) = f(x) + r(x),
    \label{objective}
\end{equation}
where $f(x)$ can be considered the negative log-likelihood and $r(x)$ the negative log-prior. It is sometimes more convenient to partition the problem such that $f(x)$ is the smooth part of the objective function and $r(x)$ the non-smooth component. The problem can then be tackled using proximal forward-backward optimisation techniques in which the forward step minimises the smooth part of the objective function $f(\cdot)$ and the backward step projects this solution onto the set that is compatible with constraints encoded in $r(\cdot)$. In particular, \citet{pfb3} investigates the convergence of a PFB algorithm when $f(\cdot)$ is Lipschitz differentiable and $r(\cdot)$ can be written as a sum of composite functions of the form
\begin{equation}
    r(x) = \sum_p \phi_p(\psi_p(x)),
    \label{regreq}
\end{equation}
where each $\phi_p(\cdot)$ is a concave, strictly increasing and differentiable function and $\psi_p(\cdot)$ is proper, lower semi-continuous and convex. The PFB algorithm utilises forward steps of the form
\begin{equation}
    \tilde{x} = x_{k} - \gamma_k U^{-1}_k \nabla f(x_k),
    \label{forward}
\end{equation}
where $0 < \gamma_k < 1$ controls the step size of the current update and $U_k$ is a Hermitian positive definite linear operator satisfying \citep[see e.g.][]{mmtut}
\begin{equation}
    f(x_{k+1}) \leq f(x_k) + (x_{k+1} - x_k)^\dagger \nabla f(x_k) + \frac{1}{2} \|x_{k+1} - x_k \|^2_{U_k}
    \label{majineq}
\end{equation}
where $\| x \|^2_U = x^\dagger U x$ is the norm of $x$ induced by the metric $U$. \citet{pfb3} show that convergence to a critical point of $\Phi(\cdot)$ can be guaranteed by using backward steps of the form
\begin{equation}
    x_{k+1} = \mbox{prox}^{U_k}_{\gamma_k q_k}(\tilde{x}) = \underset{x}{\mbox{argmin}} ~ q_k(x) + \frac{1}{2\gamma_k} \|\tilde{x} - x\|^2_U,
    \label{backward}
\end{equation}
where $q_k = \sum_p \tau_{p,k} \psi_p$ with $\tau_{p,k}>0$ weights obtained from the derivative of $\phi_p$ i.e. by majorising $r(x_k)$ using a first order Taylor expansion\footnote{Since $\phi_p$ is concave, it can be approximated by an affine function. These gives us a weighted version of $\psi_p$]}. Since the focus of the current work is on using the above framework to design practical radio interferometric imaging algorithms, the exact details will not concern us here. What is important to note is that a wide variety of optimisation problems can be written as \eqref{objective} with regularisation satisfying \eqref{regreq}, including the sparsity based method described in $\S$~\ref{sec:deconv}. Also, the proposed PFB algorithm does not require either \eqref{forward} or \eqref{backward} to have a closed form solutions so they can be solved approximately using iterative methods. The point of preconditioning is to allow using larger step sizes thus minimising the number of full gradient evaluations of $f(\cdot)$ required for convergence. This is preferable when gradient computations are expensive compared to the cost of inverting the preconditioner in \eqref{forward} and solving \eqref{backward}. Efficient strategies for solving these two sub-problems, as well as selecting algorithmic parameters, often depends on the exact form of \eqref{objective}. Thus, we henceforth specialise the discussion to the form relevant to the current work. Note, however, that there are multiple reasons (in addition to dealing with non-smooth regularisers) why one might opt to implement regularisation during the backward step. For example, it may be possible to exploit different kinds of parallelism and/or hardware during forward and backward steps to make optimal use of computing resources.

\subsection{Preconditioning for radio interferometric imaging}
\label{sec:precond}
Since \eqref{imagelik} is linear, its Hessian\footnote{The Hessian is defined as the curvature of $f(\cdot)$.} is given by
\begin{equation}
    A = R^\dagger \Sigma^{-1} R,
    \label{hessA}
\end{equation}
independent of the iteration $k$. Notice that the choice $U_k = A$ results in Newton iterations for the forward step and would therefore imbue it with second order convergence properties. This choice makes it possible to use $\gamma_k$ close to one and should therefore allow the problem to converge in relatively few forward-backward iterations (henceforth referred to as major iterations). In the absence of preconditioning, i.e. when $U_k$ is proportional to the identity $\mathrm{I}$, \eqref{majineq} dictates using $U_k = L \mathrm{I}$, where $L$ is the Lipschitz constant of $f(\cdot)$. In this case, forward steps correspond to steepest descent iterations with step size $\frac{\gamma}{L}$. Since $L$ is proportional to the largest eigenvalue of $A$, our discussion in $\S$~\ref{sec:imaging} relating the eigenvalues of $A$ to the sum of the weights on the grid now provides some intuition into why using naturally weighted gradients without any preconditioning results in algorithms with poor convergence properties. Unfortunately, the ill-posed nature of the problem implies that $A$ is not necessarily invertible.

A common strategy \citep[see e.g.][]{nocedal06} is to enforce positive definiteness by adding a small multiple of the identity to the Hessian, suggesting a preconditioner of the form
\begin{equation}
    U_R = A + \eta \mathrm{I},
    \label{precondR}
\end{equation}
where $\eta > 0$ is a constant such that $\mbox{det}|U_R| > 0$ with $\mbox{det}|\cdot|$ denoting the determinant. Unfortunately, since applying $U_R$ costs almost the same as a full gradient evaluation, this does not result in a very efficient preconditioning strategy. The computational cost can be significantly reduced by aggressive (possibly baseline dependent) averaging of the weights (i.e. $\Sigma^{-1}$ in \eqref{hessA}) and, since the accuracy of the gridding implementation we employ is tunable (see $\S$~\ref{sec:grid}), by reducing the accuracy with which $R$ and $R^\dagger$ are applied. This is one option that is exposed in the \pfb\ software package discussed in $\S$~\ref{sec:software}.

Another option is to use the approximation \eqref{hessapprox} to derive a preconditioner. However, as discussed in $\S$~\ref{sec:imaging}, the fact that $\hat{I}^{\mathrm{PSF}}$ can contain both negative and imaginary values means that a preconditioner based on \eqref{hessapprox} directly would not be positive definite or Hermitian. This can be remedied by using
\begin{equation}
    U_Z = Z^\dagger F^\dagger \hat{I} F Z  + \eta \mathrm{I},
    \label{precondZ}
\end{equation}
where $\hat{I}$ is a diagonal operator containing the absolute value of $\hat{I}^{\mathrm{PSF}}$ on its diagonal. The presence of the absolute value might seem surprising and we do not have a good theoretical justification for it at this stage. It does, however, conform to the notion that the denominator in \eqref{xpsfinvd} should correspond to the gridded weights. We will simply note that \eqref{precondZ} seems to work well in practice and defer further theoretical justifications to future work.

An important consideration is the selection of $\gamma_k$ and $\eta$. While backtracking can be used to select $\gamma_k$ such that \eqref{majineq} holds at each iteration, doing so incurs additional computational costs. It can therefore be beneficial to find an upper bound $\gamma$ on $\gamma_k$ which respects \eqref{majineq} for all iterations. The upper bound will, in general, depend on factors such as the field of view and $uv$-coverage. However, we found that $\gamma = 0.99$ works well for all the results presented in $\S$~\ref{sec:results}. This is likely due to the simple linear and convex form of $f(\cdot)$. More elaborate parametrisations might require a more careful treatment of this parameter.

The value of $\eta$ can be interpreted as the inverse variance of a zero mean Gaussian random field prior. Accordingly, larger values of $\eta$ tend to dampen the updates in \eqref{forward} while smaller values allow for a better fit to the smooth term in \eqref{objective}. In the limit as $\eta \to 0$, when using $U_R$ as the preconditioner, updates tend to the sampling density corrected gradient of $f(\cdot)$ defined as
\begin{equation}
    \delta = A^{-1} \nabla f(x_k).
    \label{natgrad}
\end{equation}
Without any additional regularisation, the sampling density corrected gradient will fit the data fidelity term of linear ill-posed inverse problems perfectly as long as the measurement model is appropriate (i.e. as long as $R$ in \eqref{immeasop} is discretised finely enough and there are no unmodelled systematics in the data). In this sense, solutions to the forward step computed using $U_R$ can be made to fit the data almost arbitrarily well by making $\eta$ sufficiently small. A similar statement holds for $U_Z$ in the coplanar array limit. Outside of this limit, since the PSF is only correct for sources at the image center, errors tend to grow with increasing distance from the center and will be proportional to the brightest unmodelled structures in the residual image. These errors dissipate towards later iterations as more structures are incorporated into the model.

The preconditioners introduced above are useful in different scenarios. The user is ultimately tasked with selecting the appropriate version and specifying a suitable value of $\eta$. We used \eqref{precondZ} with a value of $\eta = 10^{-4}$ for all results presented in $\S$~\ref{sec:results}. Such a small value allows the forward step to completely overfit the data so that updates computed using \eqref{forward} tend to an estimate of the noise in image space towards the end of the optimisation routine. Importantly, the resulting estimate is in the same units as the model image. This turns out to be useful for the kind of regularisation considered in this work (see \eqref{sigmarms}).

Finally, since the preconditioners introduced above are usually too large to compute (nevermind invert) for typical radio interferometric imaging problems, these operators are always inverted implicitly using the conjugate gradient algorithm \citep[see e.g.][]{nocedal06} which only requires the action of the operator on a vector.

\subsection{Sparsity based deconvolution}
\label{sec:deconv}
The specific form that the backward step takes depends on the regularisation. Although the framework proposed in this paper is largely agnostic of the form of regularisation employed, we specialise the discussion here to the specific form of sparsity promoting prior that is currently implemented in the \pfb\ software package discussed in $\S$~\ref{sec:software}.

The log-sum penalty function (\citep[see][]{candes07}) has been shown to be effective at promoting sparsity and has been successfully employed in interferometric imaging \cite{terris2022,thouvenin2023,wilber2023}. We utilise a variant of this regulariser to promote sparsity of $x$ in some over-complete dictionary of functions $\Psi$. Specifically, we use a regulariser of the form
\begin{equation}
    r(x) = r_+(x) + r_1(x),
\end{equation}
with $r_+(x)$ enforcing positivity and
\begin{equation}
    r_1(x) = \lambda \sum^{n_\alpha}_i \log\left( (1 + \rho_{\textrm{rms}}) \left( 1 + \frac{| g(\alpha_i) |}{\sigma_{\textrm{rms}}} \right) \right),
    \label{logsum}
\end{equation}
where $| \cdot |$ denotes the absolute value and $\alpha = \Psi^\dagger x \in \mathbb{R}^{n_b \times n_\alpha}$ is the sought after signal decomposed into the over-complete dictionary $\Psi$ that consists of $n_b$ identical copies (i.e. one for each imaging band) of $\Psi_b$ where
\begin{equation}
    \Psi_b \colon \mathbb{R}^{n_\alpha} \to \mathbb{R}^{n_p}.
\end{equation}
Here $n_\alpha$ is the total number of coefficients which is proportional to the number of pixels in the image $n_p$ and the number of bases incorporated into the over-complete dictionary. Departing slightly from the form employed in \cite{thouvenin2023,wilber2023}, we have introduced, in addition to $\lambda > 0$ which controls the overall strength of the regularisation, two hyper-parameters viz. $\rho_{\textrm{rms}} > 0$ and $\sigma_{\textrm{rms}} > 0$. As discussed below, the problem can be coerced into a form where these parameters take on an intuitive meaning. The choice of $\Psi$ considered in this work is a concatenation of the identity and up to the first eight Daubechies wavelets. This form of $\Psi$ has been extensively used in the SARA family of algorithms first introduced in \cite{sara}.

The function $g : \mathbb{R}^{n_b} \to \mathbb{R}$ is a function used to further regularise the problem along the frequency axis, typically chosen to be some kind of norm (with the Euclidean norm being a popular choice). We have found that
\begin{equation}
    g(\alpha_i) = \sum_b^{n_b} \alpha_{b,i},
    \label{gfunc}
\end{equation}
performs better for continuum imaging. This choice of function enforces sparsity of the mean image\footnote{It is worth noting that we do not normalise the residual image by the sum of the weights, routinely referred to as $w_{\mathrm{sum}}$, that is implied by the $R^\dagger W$ operator in a per band manner. This would over/under-emphasise bands with small/large $w_{\mathrm{sum}}$ values.}.

The function \eqref{logsum} adheres to the form \eqref{regreq}. Indeed, \citet{pfb3} contains an explicit example of using log-sum regularisers within the optimisation framework detailed in $\S$~\ref{sec:pfb}. Since \eqref{logsum} is not a convex function, care has to be taken with the initialisation. By interpreting the problem in a majorise-minimise framework, log-sum regularisers can be implemented as a sequence of re-weighted L1 regularisers. For the specific form \eqref{logsum}, each iteration of such a sequence uses a convex regulariser of the form
\begin{equation}
    q_{k,\tau}(x_k) = \lambda \sum_i^{n_\alpha} \tau_{i,k} | g(\alpha_{i,k}) |,
    \label{reg21W}
\end{equation}
with the $\tau_{i,k}$, also referred to as L1-weights, initialised to unity. As long as $\prox_{q_{k,\tau}}(\cdot)$ is tractable, the solution to \eqref{backward} can be found approximately using the primal-dual algorithm (see \cite{condat2012} for example). The $\tau_i$ are updated at convergence of each major iteration of the algorithm according to
\begin{equation}
    \tau_{i,k} = \frac{1 + \rho_{\textrm{rms}}}{1 + \frac{| g(\alpha_{i,k}) |}{\sigma_{\textrm{rms}}}},
    \label{rwgt}
\end{equation}
which we refer to as L1-reweighting. Note that, if $\sigma_{\textrm{rms}}$ is an estimate of the standard deviation of the noise projected into $g(\Psi^\dagger(\cdot))$ space, then $\rho_{\textrm{rms}}$ can be set to determine how aggressively the L1-reweighting should proceed. For instance, setting $\rho_{\textrm{rms}} = 1$ results in unity $\tau_i$ for model components with an signal to noise ratio (SNR) of one\footnote{The definition of SNR employed here is simply $\frac{x}{\mbox{std}(x)}$ where $\mbox{std}(\cdot)$ denotes the standard deviation of its argument.}. Clearly, components with an SNR greater/smaller than one experience smaller/larger L1-weights respectively. Setting $\rho_{\textrm{rms}} > 1$ raises the level of SNR for which unity L1-weights are assigned. This can be useful during the early stages of the reduction to suppress calibration artefacts and/or artefacts stemming from unflagged RFI. Of course, the key requirement for \eqref{rwgt} to function as intended is for $\frac{|g(\alpha_i)|}{\sigma_{\textrm{rms}}}$ to evaluate to the SNR of component $\alpha_i$. This can be achieved in an automated way by utilising properties of the sampling density corrected gradient \eqref{natgrad} at the current iteration.

As suggested by the formula \eqref{rwgt}, the key to automating hyper-parameter selection is understanding the units that the problem is expressed in. Since the discrete form of the integral operator \eqref{RIME} has a volume factor associated with it (i.e. the area of a pixel) the numerical vector $x$ in \eqref{discRIME} has units of Jy/pixel. The residual image, on the other hand, has the same units as $A x$. Clearly, some form of unit conversion is required if $\sigma_{\textrm{rms}}$ is to be determined from the noise level while simultaneously having the ratio $\frac{g(\alpha_i)}{\sigma_{\textrm{rms}}}$ evaluate to an SNR. The sampling density corrected gradient \eqref{natgrad} accounts for this conversion and can, in principle, be used to compute $\sigma_{\textrm{rms}}$ as
\begin{equation}
    \sigma_{\textrm{rms}} = \mbox{std}(\alpha_\delta), \qquad \alpha_{\delta,i} = g([\Psi^\dagger \delta]_i),
    \label{sigmarms}
\end{equation}
with $\delta$ as defined in \eqref{natgrad} and where $\mbox{std}(\cdot)$ denotes standard deviation. In practice, since $A$ is not directly invertible, we estimate $\sigma_{\textrm{rms}}$ by substituting $\tilde{\delta} = U^{-1} I^R \approx \delta$ into \eqref{sigmarms} where $U$ is the preconditioning operator. Since the residual only tends to be noise like towards the end of the optimisation algorithm, prematurely triggering L1-reweighting can preclude structures from getting into the model. It is therefore advisable to let the algorithm run to convergence, or for a fixed number of major iterations, before triggering L1-reweighting. Once triggered, we re-evaluate $\sigma_{\textrm{rms}}$ and recompute the L1-weights using \eqref{rwgt} after each gradient evaluation.

\begin{figure*}
    \centering
    \includegraphics[width=2\columnwidth]{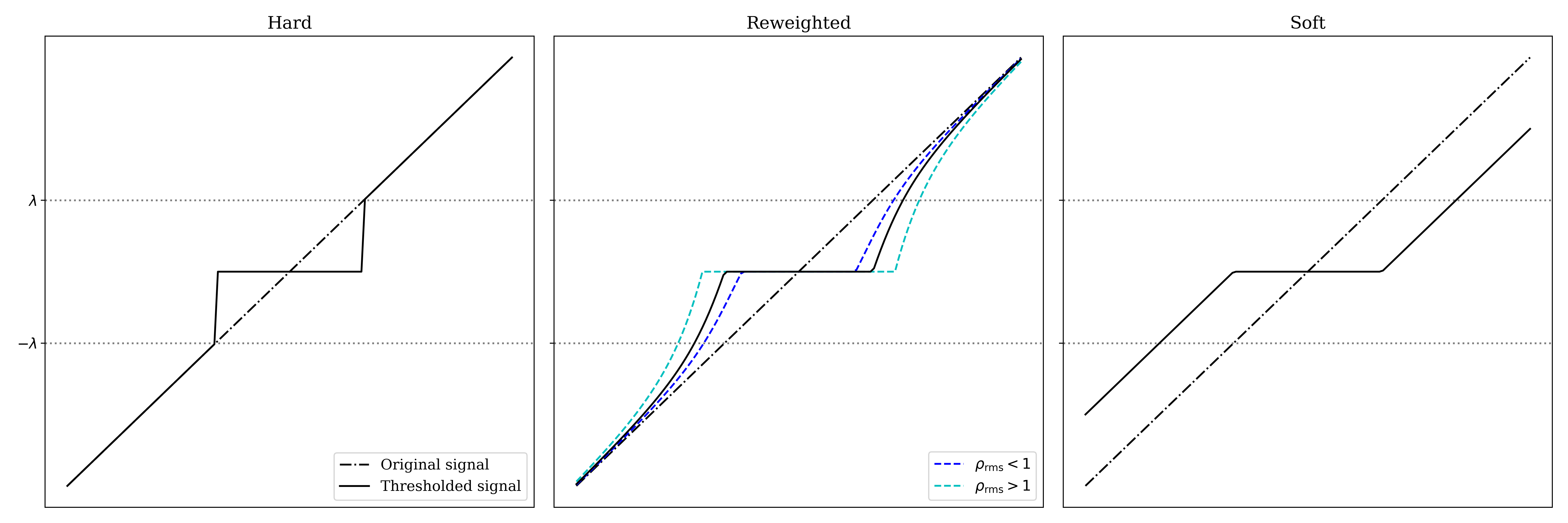}
    \caption{Different form of thresholding. Left - hard thresholding. Right - soft thresholding. Middle - the thresholding operation corresponding to minimising the re-weighted L1 norm in \eqref{reg21W}. The latter operation is a trade-off between hard and soft thresholding that sets noise-like components to zero while allowing bright components to be thresholded relative to their SNR. The dotted horizontal line corresponds to the overall level of thresholding that is applied for a given value of the parameter $\lambda$. The dashed coloured lines in the central figure illustrate the effect of the parameter $\rho_{\textrm{rms}}$. }
    \label{fig:thresholding}
\end{figure*}

Next, we turn to the parameter $\lambda$. Sparsity based methods inevitably use some sort of soft-thresholding operation to implement the proximal operator corresponding to \eqref{reg21W}. Different forms of thresholding are illustrated in Figure~\ref{fig:thresholding}. Note, in particular, the form of the thresholding operation corresponding to minimising the re-weighted L1 norm, shown in the middle panel. This operation sets all components below a certain value to zero and decreases larger components in a way that is inversely proportional to their magnitude. If the value of $\lambda$ is chosen appropriately, this acts as a soft ``mask" which thresholds components relative to their SNR, with noise-like components being set to zero. This suggests that an estimate of the noise in the space where thresholding happens can be used to set this parameter in an intuitive way. In our case, noise enters the backward step via the smooth part of the objective function in \eqref{backward} or, more accurately, its gradient which is proportional to $U(\tilde{x}-x)$. In the limit as $k \to \infty$ we can substitute for the $\tilde{x} - x$ from \eqref{forward} to obtain
\begin{equation}
    U(\tilde{x}-x_k) = -\gamma_k I^R,
\end{equation}
showing that the noise in the backward step remains proportional to the residual image, even in the presence of preconditioning. We therefore set the thresholding parameter $\lambda$ in \eqref{reg21W} according to
\begin{equation}
    \lambda = \rho_{\textrm{rms}} \mbox{std}(\alpha_R), \qquad \alpha_{R,i} = g([\Psi^\dagger I^R]_i),
    \label{setlam}
\end{equation}
where $\rho_{\textrm{rms}}$ is the same as in \eqref{rwgt} and $I^R$ is the residual image at the current major iteration. Practically this means that $\lambda$ starts out quite large and decreases down to a limiting value that will be related to the rms of the noise in image space. While this strategy does not lead to the most rapid convergence, it does tend to stabilise the algorithm and prevent artefacts from entering the model, especially if L1-reweighting is triggered timeously. The value of $\lambda$ can also be specified manually which is sometimes useful but we have found \eqref{setlam} to be very effective way to initialise this parameter in practice. It also means that hyper-parameter specification, at least at the outset, boils down to setting the single parameter $\rho_{\textrm{rms}}$ which has an intuitive interpretation in terms of the level of thresholding that should be applied as explained above. There are a number of additional algorithmic parameters that need to be set for the primal dual algorithm (e.g. the primal and dual step sizes, convergence criteria etc.) but these mainly affect the rate of convergence of the algorithm. A lot of effort has gone into providing sensible defaults but some of the parameters tend to be observation specific. This is discussed further in $\S$~\ref{sec:sara}.

Finally, the positivity constraint $r_+(x)$, when present, can take one of two forms. In its usual form it simply amounts to setting negative model components to zero. A stronger form can also be imposed to further suppress artefacts and improve point source reconstruction. In this form, components which are less than zero in any imaging band are set to zero. This is a rather strict form of regularisation which is mainly useful during the early stages of self-calibration.

To summarise, the preconditioners detailed in $\S$~\ref{sec:precond} make it possible to use a larger step size in the PFB algorithm by accounting for the approximate local curvature of the data fidelity term $f(\cdot)$ during the forward step \eqref{forward}. In the context of radio interferometric imaging, this essentially amounts to correcting the updates for the sampling density of the interferometer or, in other words, giving all spatial scales equal emphasis. The backward step then imposes regularisation while accounting for the approximate uncertainty in each spatial scale, as depicted in Figure~\ref{fig:weighting}, by incorporating an estimate of the local curvature (i.e. $U$) into the smooth part of \eqref{backward}. This allows the algorithm to utilise more naturally weighted data without sacrificing resolution. However, even though a larger overall step size (i.e. $\gamma$) can be used, both the forward and backward steps will exhibit slower convergence for more naturally weighted data. In the case of the forward step, this happens because the rate of convergence of the conjugate gradient algorithm is inversely proportional to the conditioning number of $U$ (i.e. the ratio of its largest to the smallest eigenvalues) which is larger for naturally weighted data. Similarly, during the backward step, the step size used in the primal dual is inversely proportional to the largest eigenvalue of $U$. Visibility weighting schemes can therefore be used to further accelerate convergence but this comes at the cost of reduced sensitivity. With all the algorithmic components in place, we now turn to the practical implementation.

\section{The \pfb\ Software Package}
\label{sec:software}
This section provides details about the \pfb\ software package\footnote{\url{https://github.com/ratt-ru/pfb-imaging}} which implements the imaging framework described in this paper. We'll start with an overview of the design philosophy and technologies that are employed and then move on to the specific applications that are available.

\subsection{Overview}
In keeping with the philosophy of the current paper series \cite{africanus1,africanus2,africanus4}, \pfb\ attempts to leverage technologies that are maintained by a large and active open source community as far as possible. Since these technologies are described in detail in the accompanying papers of this series, they are only briefly discussed here.

For ease of development, \pfb\ is written mainly in \python\ \cite{vanrossum2009}. Being an interpreted language, \python\ is known to be slow compared to compiled languages like C, C++ and Fortran. \pfb\ therefore uses \numba\ \cite{lam2015} to just-in-time (JIT) compile performance critical components for which high performance extensions are not available (via e.g. \numpy\ \cite{Harris2020} and/or \scipy\ \cite{Virtanen2020}). This also makes it possible to bypass \python's global interpreter lock when executing certain tasks in parallel. Operations on data that can be partitioned into independent atomic units (i.e. chunks) are parallelised and optionally distributed using \dask\ \cite{rocklin2015}. In addition, a form of nested parallelism is available to give users more control over resource utilisation. In essence, users can choose to process as many chunks (often imaging bands) as can fit in memory in parallel and then further parallelise compute intensive tasks like gridding, FFTs and wavelet transforms within a chunk. Such tasks can often be parallelised with minimal memory overhead.

The previous two papers in the series \cite{africanus1,africanus2} illustrate the scaling of the \dask\ distributed scheduler using its collections interface to solve embarrassingly parallel problems. It is shown that the scheduler can be coerced (e.g. by carefully cloning root nodes and/or using scheduler plugins) to scale linearly with the amount of resources allocated as long as the problem can be partitioned into a sufficient number of (large enough) chunks. However, the need for regularisation when solving ill-posed inverse problems can make it impossible to partition algorithms in an embarrassingly parallel way and, in general, different forms of regularisation will require different communication patterns between workers. We have observed that the distributed scheduler often fails to respect data locality when attempting to distribute such problems using the collections interface, especially when attempting to embed multiple fan-reduce type operations within a graph. Our attempts to wrangle the collections interface into a form that does not result in unnecessary data transfers between nodes also resulted in unnatural looking code, often obscuring the underlying algorithm.

One of the main aims driving the development of \pfb\ is to create a flexible and developer friendly environment in which to foster new algorithms. It therefore departs, in most places, from the more rigid declarative programming style necessitated by the collections interface and uses Dask's client interface instead. While this sacrifices resilience to cluster failures, it allows for a more imperative programming style based on futures which should be more familiar to developers. All data products produced by \pfb\ use the \xarray\ \cite{hoyer2017} format and are backed by \zarr{}\footnote{\url{https://zarr.dev/}} \cite{Zarr2018} on disk. Individual chunks are stored as separate \xarray\ {\tt DataSet}s in the same directory store. This allows for a simple lazy loading mechanism which does not require \dask\ collections. The collections interface is only invoked when interfacing with raw interferometric data.

\pfb\ interfaces with interferometric data using \daskms\ \cite{africanus1}. This makes it agnostic to data sources in the form of CASA \cite{casa2022} Table Data System (CTDS) \citep{vandiepen2015}, \zarr\ or \textsc{Parquet}\footnote{\url{https://parquet.apache.org/}} \cite{Arrow2019} backed measurement sets. The latter two formats allow for efficient parallel read and write operations and are natively compatible with object storage. This opens the door to commodity computing platforms such as \aws\ while retaining the ability to directly process legacy CTDS backed measurement sets. Images can optionally be rendered to FITS \cite{Wells1981} but special storage has to be provisioned when using computing facilities that rely on object storage since FITS is not natively compatible. The metadata required to produce FITS headers are stored as attributes of the \xarray\ dataset and a dedicated application is available to convert images to FITS files.

In an attempt to keep the implementation as flexible as possible, \pfb\ comprises a number of distinct applications, each with their own command line interface (CLI). CLIs are constructed using the \click\footnote{\url{https://click.palletsprojects.com/}} package by using the \texttt{clickify\_parameters} decorator available in \stimela\ (see $\S$~5.5 of \cite{africanus4}). Applications are grouped under a command group called \texttt{pfb} which is added as an executable during installation. Each application can be run individually from the terminal or grouped together into a highly customisable imager with \stimela. The cab definitions (see $\S$~4 in \cite{africanus4}) for all applications are defined in \pfb{}, obviating the need for users to provide them manually. \stimela\ can also be used to distribute the imager over a computing cluster. When running in a \kubernetes\ \cite{brewer2015} enabled distributed environment, \stimela\ takes care of launching the requested \dask\ clusters and requires minimal expertise on the part of the user. Importantly, different clusters can be requested for each step in the recipe allowing for efficient use of resources even when running computationally heterogeneous recipes. This is demonstrated using a \texttt{LocalCluster} in the accompanying \stimela\ recipe in \ref{sec:recipes} used for some of the results presented in $\S$~\ref{sec:results}. The amendment to this recipe which enables deploying the same workflow on \aws\ instances is given in Appendix B of \cite{africanus4}. Interested readers are encouraged to consult \cite{africanus4} for further details.

Our deconstructed implementation of the imager serves another purpose viz. to make testing and prototyping new features easier. While this does have some impact on performance\footnote{The performance of \pfb\ is benchmarked against the popular \wsclean\ imaging package in $\S$~\ref{sec:results}.}, and requires maintaining a larger code base, we believe that the increased accessibility is more important in the long term, especially given the rapidly evolving landscape of interferometric imaging algorithms. Abstracting algorithmic details from data formats should enable rapid prototyping of new algorithms by developers who are not necessarily experts in the field.

\subsection{{\sc pfb-imaging} applications}
What follows is a breakdown of the main applications that are currently available in \pfb. Many of the practical aspects underlying its design are more easily communicated by documenting the function of the individual applications in this way. The order in which the applications are introduced also roughly matches the order in which they would be deployed in a typical imaging pipeline.

\subsubsection{Initialising Stokes visibilities}
\label{sec:init}
The \init\ application is responsible for casting data into a form more conducive to the imaging problem. In particular, given the raw data and corresponding net gain table produced by \qcal{}, or corrected data, the individual Stokes visibilities are computed according to equation \eqref{cdata} and the corresponding weights are obtained from the diagonal of \eqref{cweight}. Metadata that might be required by other steps in the pipeline (e.g. antenna labels, channel widths and integration time interval) can often be discarded during imaging making the data much more amenable to averaging. Both simple averaging in frequency and baseline dependent averaging in time\footnote{Baseline dependent averaging along the frequency axis is not currently supported as it results in ragged arrays which are not currently supported by the \wgridder\ in \ducc .} are supported and make use of the averaging modules in \codexafricanus\footnote{\url{https://github.com/ratt-ru/codex-africanus}} \cite{africanus1}. The degree of averaging is controlled by specifying the number of visibility channels to average together and the level of decorrelation that can be tolerated at some maximum radial distance from the field center (via the \textsf{max-field-of-view} parameter).

In order to respect data discontinuities, the data are always partitioned by scan, spectral window and field (per measurement set) at the outset. The temporal and spectral resolution can be further refined to a user specified number of time integrations and channels per image. Since BDA potentially results in a different number of output rows when averaging over different frequency ranges, the partitioned data can't always be combined into a regular array and are therefore written to separate \xarray\ datasets, along with all relevant metadata. This partitioning by dataset, which is trivial to parallelise over the time and frequency axes, also enables a simple lazy loading (i.e. only loading data upon request) mechanism using \zarr. Since the throughput of object stores such as Amazon S3 is anticipated to scale linearly with the number of simultaneously connected workers, provided they have independent network connections, it is possible to get good throughput from S3 by using many workers for this task.

The \init\ application provides a simple data selection mechanism that can be used to image a subset of input data. At the time of writing, it is possible to select data by field, spectral window, scan and frequency range per measurement set. It is currently assumed that all fields selected for imaging share a common phase center but in principle multiple fields can be rephased to a common phase center at this stage. Average primary beam patterns that are suitable for the time and frequency ranges in each dataset can be computed if antenna primary beam patterns are available for the array. This, along with further refinements to the available selection mechanisms, is deferred to future work.

\subsubsection{Initialising image data products}
\label{sec:grid}
Before any actual deconvolution can be performed, the outputs of the \init\ application need to be projected into a suitably defined image space. The \grid\ application is responsible for setting parameters related to the discretised measurement operator \eqref{immeasop} and can be used to compute image space data products such as the dirty, PSF and residual images. Since users may need to change imaging parameters such as pixel sizes, output dimensions (e.g. number of bands and pixels) or the image center, it is beneficial to cache the outputs of the \grid\ application separately from those of the \init\ application. This also makes it possible to image multiple objects of interest (e.g. the sun and the main field) simultaneously.

The \grid\ application also computes imaging weights using the standard Briggs weighting scheme \cite{briggs1995high} by exposing a single \textsf{robustness} parameter. The imaging weights and $uvw$ coordinates are written alongside the image data products. The datasets produced by the \grid\ application are ingested and modified by the deconvolution applications discussed in the next section, effectively serving as the state of the imager at any given time. These applications avoid the need to load visibilities into memory by computing residuals using \eqref{visgrad}. This reduces the memory footprint but also means that the deconvolution applications don't have direct access to the visibilities and therefore can't, for example, change the robustness value of the Briggs weighting scheme or do any flagging based on outliers directly. The \grid\ application has to be rerun for such operations. However, it does attempt to automatically detect cases in which cached data products can be reused by examining dataset metadata.

All gridding operations are currently performed using the improved w-stacking algorithm \cite{Ye_2021} implemented in the \wgridder\ \cite{wgridder} in \ducc\footnote{\url{https://gitlab.mpcdf.mpg.de/mtr/ducc}}. The \wgridder\ parallelises natively with low memory overheads but its efficiency\footnote{Efficiency refers to single-threaded walltime divided by nthreads times multi-threaded walltime.} degrades with increasing thread count so nested parallelism (e.g. gridding multiple bands simultaneously) is beneficial. The \wgridder\ also allows tuning the gridding accuracy by setting the maximum L2-error compared to the direct Fourier transform. This can be used to reduce the cost of applying the operator $U_R$ in \eqref{precondR}. Since the \wgridder\ is also available in \wsclean{}, the comparisons we present in $\S$~\ref{sec:results} utilise the same measurement operator.

\subsubsection{Deconvolution}
\label{sec:sara}
There are multiple deconvolution algorithms, including a variant of the \clean\ algorithm, implemented in \pfb. All of them use the same data structures, produce the same data products and are based on the theory detailed in $\S$~\ref{sec:measmod} and $\S$~\ref{sec:inverse}. In particular, they all interface with the data products produced by the \grid\ application which are used to store the state of the algorithm after each major iteration. The homogeneous deconvolution interface makes it possible to chain these algorithms together so that one can pick up where another left off (an example of this is described below). This section only details the two deconvolution applications used to produce the results presented in $\S$~\ref{sec:results}.

{\tt sara}: The most mature deconvolution algorithm currently in \pfb\ is invoked by the \sara\ application. This is the algorithm described in $\S$~\ref{sec:deconv}. There are essentially three optimisation routines in the \sara\ application for which algorithmic parameters need to be specified, in addition to the outer preconditioned forward-backward algorithm. These are:
\begin{itemize}
    \item the \emph{power method} to approximate the spectral norm of the preconditioner,
    \item the \emph{conjugate gradient} algorithm used to invert the preconditioner in \eqref{forward} when an explicit inverse is not available and,
    \item the \emph{primal-dual} algorithm used to find the solution to the backward step \eqref{backward}.
\end{itemize}
Since the algorithmic parameters (e.g. stopping criteria, maximum number of iterations, step sizes etc.) need to be tuned for different kinds of observations, these are exposed in the CLIs for each application. While a great deal of effort has gone into setting sensible defaults, and to give the hyper-parameters an intuitive meaning as explained in $\S$~\ref{sec:deconv}, the onus is on the user to set these parameters for their specific observation. Note that model and residual images can be written to disk after each iteration to help monitor the progress of the algorithm which can safely be interrupted at any stage. Caching the state of the deconvolution algorithm at every iteration also facilitates manual inspection and monitoring which should hopefully make tuning algorithmic parameters easier. Finally, one of the benefits of encapsulating applications in the form of \stimela\ recipes is that it allows for exposing only a subset of relevant parameters. Once a particular recipe is tuned for a specific kind of observation, users should be able to apply it to similar observations without needing to set all the algorithmic parameters manually.

The \sara\ application relies purely on the native parallelisation of the operators (e.g. gridding, FFT and wavelet transforms) involved and is not currently distributed in any way. This keeps it as simple and memory efficient as possible and provides a reference that distributed implementations can be verified against. Although a prototype implementation of \sara\ that is distributed over the frequency axis is in principle available, the kind of regularisation employed means that naively distributing over frequency necessitates communicating large dense arrays between nodes. We have observed that this introduces communication overheads which, for problems of the size considered in $\S$~\ref{sec:results}, can be comparable to the time spent executing the parallelisable parts of the algorithm. We therefore defer a better distribution strategy to future work. Note that, because of averaging and preconditioning, the computational complexity and memory requirements of \sara\ are often dominated by operations involving the regulariser which scale with the size of the image and not that of the data. This can, however, be quite significant because of the need to compute wavelet transforms for multiple wavelet bases.

\pfb\ contains a custom parallel \numba\ implementation of 2D discrete wavelet transforms (DWT) for the Daubechies \cite{Daubechies1988} family (filter bank coefficients are obtained from \textsc{PyWavelets} \cite{Lee2019}). Our implementation uses \numba{}'s \texttt{prange} to perform multiple 1D transforms (corresponding to rows of the input signal) per basis in parallel. This allows for parallelism with only minor memory overheads. Since the entire 2D function call is JIT compiled, it is also possible to parallelise over bases using standard \python\ threads but this introduces additional memory pressure. As we'll see in $\S$~\ref{sec:results}, the need to hold multiple bases in memory often limits the kind of hardware that \sara\ can be deployed on.

{\tt fluxtractor}: The more expressive model format afforded by the wavelet based prior necessitates the artefact mitigation strategies discussed in $\S$~\ref{sec:deconv} (e.g. the strong form of the positivity constraint), especially at the early stages of the reduction. Even with these strategies in place, deconvolution depth is often limited by the presence of calibration artefacts. \clean\ makes use of masking strategies to maximise model completeness without incorporating calibration artefacts into the model. \pfb\ implements an efficient way to extract all the flux within a given mask via the \fluxmop\ application. This application uses the conjugate gradient algorithm to compute $\tilde{x}$ in \eqref{forward} with the chosen preconditioner while optionally imposing a mask. The resulting model can contain unphysical features such as negative components but it will usually be a very good fit to the data. As with \clean{}, the quality of the result often depends strongly on the input mask.

The \fluxmop\ application is useful for different reasons at different stages of the reduction. One such stage is immediately preceding the first round of self-calibration since, given a good quality mask, it will extract all the model flux in unmasked locations. It can also be useful when doing image space searches for transients by looking at high time cadence residuals \citep[see e.g.][]{smirnov2024}), or when looking for spectral line emission. In these cases, it is not the mopped models but the noise-like residuals that accompany them that are scientifically interesting. The residuals can also be indicative of remaining unmodelled systematics in the data (e.g. direction dependent effects) since it is usually impossible to fit the data perfectly using the discretised form of the measurement operator \eqref{immeasop} when such systematics are present.

\subsubsection{Degridding and the model format}
\label{sec:degrid}
In addition to model images in the form of regular arrays, \pfb\ exposes a model format which supports parametrisation in terms of arbitrary continuous functions. This is achieved by keeping track of all non-zero spatial pixels and fitting a parameterized function to the frequency axis of each pixel independently. The fitted coefficients and corresponding locations are written to a single \xarray\ dataset with the parametrisation encoded in its attributes. The resulting expression can be used to evaluate the model at the original spatial resolution at any frequency. An option is exposed to allow using the sum of the weights in each imaging band as weights during the fit to the frequency axis, thus providing some degree of smoothing to avoid biases that may stem from heavily flagged imaging bands.

The resolution of the spatial axes can also be changed to some extent by using the efficient \texttt{RegularGridInterpolator} routines available in \scipy. Since the model is in units of Jy/pixel, the flux in output pixels needs to be scaled by the ratio of the input to output pixel areas. This form of interpolation is not perfectly flux conservative and more sophisticated flux conservative interpolators are currently being investigated (e.g. by using the non-uniform Fourier transform). Imperfect spatial interpolation does not pose a huge problem when continuing a deconvolution run with a different resolution as the algorithm will correct for the slight discrepancies introduced.

All deconvolution applications produce a model in this output format. The default parametrisation of the frequency axis consists of orthogonal Legendre polynomials with an order that matches the number of non-null input bands, resulting in near perfect interpolation within the domain of the problem. In addition, \pfb\ provides a \modeltocomps\ application to facilitate changing the model parametrisation. The intention of the model format is not solely compression but also to allow downstream applications to parse the model and render it to a grid of the desired resolution (e.g. a downstream calibration algorithm) without any ambiguity in the parametrisation that has been employed. The model specification is still highly experimental and likely to change in the future. However, a versioning scheme, along with dedicated utility functions that can be used to render the model to a grid, is in place to prevent models from becoming obsolete. An example of the current model format is shown in Figure~\ref{fig:xarray_dataset}.
\begin{figure*}
\begin{verbatim}
    <xarray.Dataset> Size: 11MB
    Dimensions:       (par: 4, comps: 220432, f: 4, x: 220432, y: 220432, t: 1)
    Coordinates:
        freqs         (f) float64 32B ...
        location_x    (x) int64 2MB ...
        location_y    (y) int64 2MB ...
        params        (par) <U2 32B ...
        times         (t) float64 8B ...
    Dimensions without coordinates: par, comps, f, x, y, t
    Data variables:
        coefficients  (par, comps) float64 7MB ...
    Attributes:
        cell_rad_x:       3.2594368590740456e-05
        cell_rad_y:       3.2594368590740456e-05
        center_x:         0.0
        center_y:         0.0
        ra:               0.0
        dec:              0.5235987755982988
        fexpr:            3.33333333333333e-9*f - 4.66666666666667
        npix_x:           810
        npix_y:           810
        parametrisation:  f*f1 + f2*(3*f**2/2 - 1/2) + f3*(5*f**3/2 - 3*f/2) + t0
        spec:             genesis
        stokes:           I
        texpr:            t
\end{verbatim}
\caption{The \xarray{} representation of \pfb{}'s model format.}
\label{fig:xarray_dataset}
\end{figure*}

\pfb\ also provides a \degrid\ application to render models in the above format to visibilities and write them to an arbitrary measurement set column using \daskms. The parametrisation of the frequency axis makes it possible to refine (or coarsen) the frequency resolution with which the model is rendered to visibilities. Work is ongoing to incorporate this degridding mechanism directly into \qcal. Since \pfb\ also knows how to parse \qcal\ gain models, this potentially circumvents the need for additional visibility sized data products. Combining this with some sort of (possibly automated) model segmentation to perform direction dependent calibration is an avenue we plan to pursue in future work (e.g. using region files to select problem sources that need to peeled).

While the \degrid\ step is also embarrassingly parallel, and there is a significant amount of work to do per chunk, the underlying storage format dictates whether chunks can be written in parallel. Nested parallelism is therefore also useful here as it allows optimising parallelism settings for the storage format in use.

\subsubsection{Image restoration}
\label{sec:restore}
Scientifically interesting parameters are not usually estimated from model images but rather from a data product referred to as the restored image. A restored image is created by fitting a Gaussian to the main lobe of the PSF, convolving the model with the result, and adding the residuals back in. The process is necessitated by the fact that model images, especially those produced by variants of the \clean\ algorithm, tend to be non-physical and, by virtue of being point estimates, do not account for any uncertainty in the reconstruction. Since a convolution with a Gaussian effectively down-weights long baselines, the very ones that are usually the least well sampled, restored images at least have some notion of positional uncertainty folded in. The residuals are added in to retain undeconvolved flux. This also makes it possible to get some kind of handle on the uncertainty in flux for different structures in the image since the signal can now be measured relative to the noise floor of the observation. Presumably for these reasons, restored images are routinely used as the data product from which scientifically interesting parameters can be derived. In deference to the status quo, \pfb\ has a \restore\ application which can be used to produce restored images to fit scientific requirements, homogenising the resolution across imaging bands if required.

It is unfortunate that astronomers still have to resort to such crude estimation methods (see \cite{Arras_2021} for a detailed discussion about the limitations of restored images and the need for reliable uncertainty estimates) when dealing with large radio interferometric data. Despite many ongoing efforts \citep[see e.g.][]{sutter2014,resolve1,Arras_2021,roth2024fastresolvefastbayesianradio,liaudat2024} we are not aware of a framework that can derive meaningful posteriors for large scale interferometric inverse problems, at least not without specialist hardware. Such a framework should ideally fold in systematic uncertainties stemming from imperfect calibration, unflagged RFI and imperfect knowledge of measurement noise. Unfortunately, \pfb\ does not offer a solution at this stage.

\section{Results}
\label{sec:results}

\subsection{Initial processing}
We validate our imaging framework on observations of the \eso\ field collected using the MeerKAT telescope (project ID SCI-20190418-SM-01). The data consists of two separate observations both spanning approximately 8 hours with 8 seconds integration time and with a spectral resolution of approximately 209 kHz in L-band (0.856 GHz-1.712 GHz) resulting in 4096 frequency channels. The primary calibrator PKS B 1934-638 is sufficiently close to the target field to serve as both bandpass and gain calibrator. Both observations consist of 15 scans where scans alternate between 6 minute observations of  PKS B 1934-638 and 1 hour observations of \eso. We used a combination of auto-flagging with \textsc{Tricolour}\footnote{\url{https://github.com/ratt-ru/tricolour}} and tedious manual inspection to flag the data. The first and last 128 frequency channels were flagged to avoid spectral roll-off at the edges of the band. Frequency channels that are known to be affected by persistent RFI were also flagged. \qcal\ \cite{africanus2} was used to produce initial gain estimates for the target field using the model \eqref{jchain} with the strategy detailed in \ref{sec:calibration}. The \init\ application was used to compute Stokes visibilities and weights averaged by a factor of 4 in frequency and using BDA with a maximum decorrelation factor of 0.98 at a field of view of $2^\circ$. The resulting dataset is approximately 100GB on disk which amounts to about a factor of 20 in compression. The same level of averaging is used throughout.

\begin{figure*}
    \centering
    \includegraphics[width=2\columnwidth]{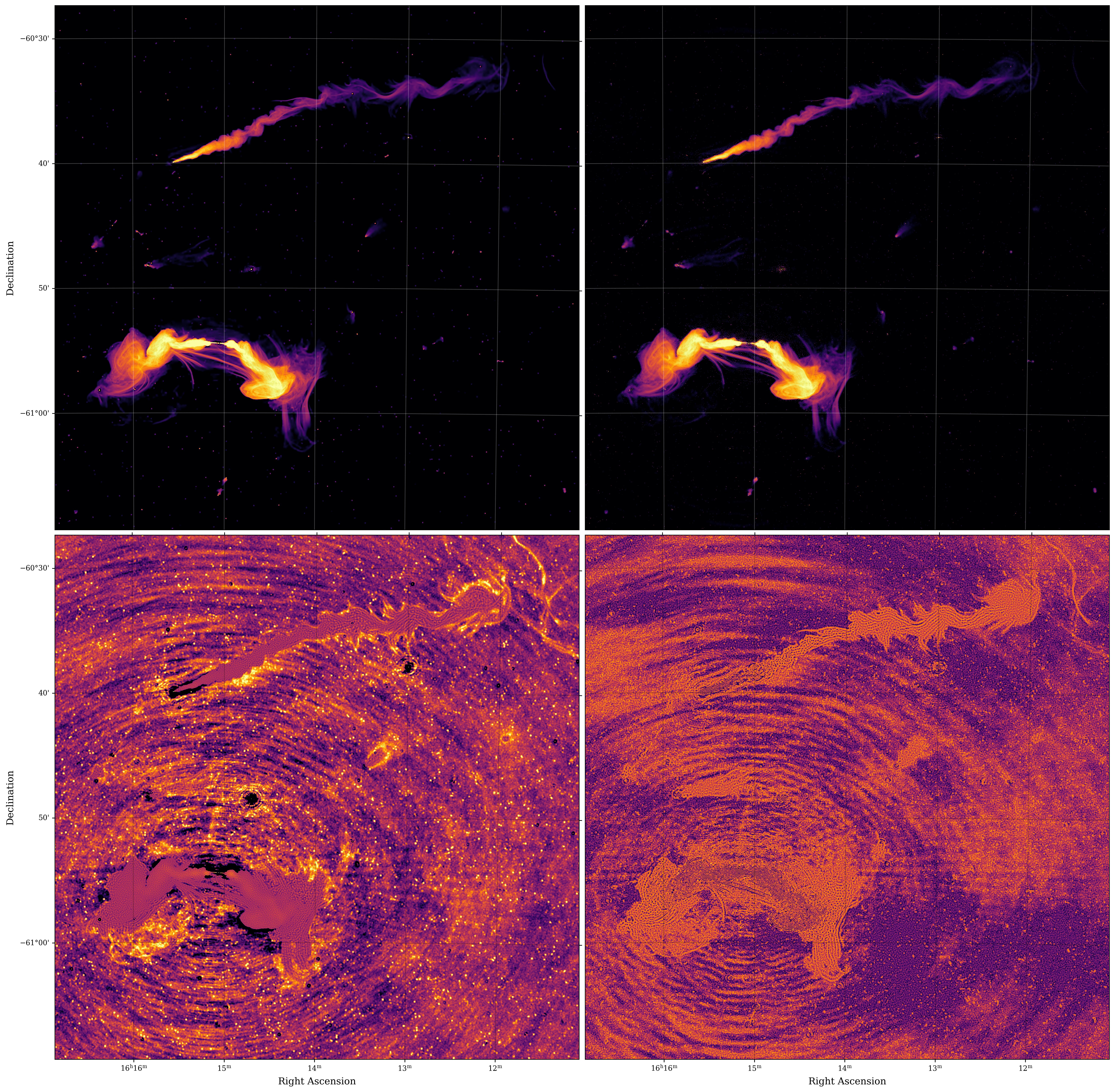}
    \caption{Comparison between \pfb\ (left) and \wsclean\ (right) on the ESO137 field after the final round of self-calibration. Top panel shows scale matched model images in log scale. The bottom panel shows scale matched residual images in linear scale.}
    \label{fig:pfbvwsc}
\end{figure*}

The \grid\ worker was then used to create the imaging data products as discussed in $\S$~\ref{sec:grid}. The image was discretised into 20 frequency bands (i.e. 192 frequency channels per band after discarding the edges) and $1^{\prime\prime} \times 1^{\prime\prime}$ pixels. This resulted in a $20\times 7580\times 7580$ image covering a field of view of approximately $2.0^\circ$. To accelerate convergence and suppress systematics affecting short baselines, we used a Briggs robustness factor of $-1$ at the outset. An initial round of deconvolution was performed using the \sara\ application with $\rho_{\textrm{rms}}$ set to 3.5 and using the strong form of the positivity constraint as discussed in $\S$~\ref{sec:deconv}. The application was terminated after 5 major iterations at which point regularisation becomes too weak to preclude artefacts from entering the model. To maximise model completeness, we created a mask using \textsc{breizorro}\footnote{\url{https://github.com/ratt-ru/breizorro}} and used the \fluxmop\ application to extract the remaining flux within the mask. The resulting model was rendered to components using the \modeltocomps\ application with the default parametrisation and number of basis functions set to one less than the number of imaging bands. The model was up-sampled to 128 channels per band and rendered to model visibilities on disk using the \degrid\ application. The model visibilities were written to the original measurement set and \qcal\ was used to calibrate for residual delay and phase using the calibration model \eqref{jchain2}. The resulting net gains were used for a subsequent round of deconvolution using the same parameters as before except this time using a robustness factor of -0.5, $\rho_{\textrm{rms}} = 2$ and using the weak form of the positivity constraint. We ran the application for 10 major iterations triggering L1-reweighting on the 5th iteration.

Although a significant amount of undeconvolved structure remains in the residuals at this stage, we found that relaxing the regularisation resulted in unphysical features being incorporated into the model. We therefore used \qcal\ with the model \eqref{jchainf} to solve for both amplitudes and phases which resulted in a visible reduction to the background noise in the residual images. To set up a fair comparison with \wsclean{}, we used \qcal\ to create the corrected data and weights using this final gain model. We then used \xova\ \cite{atemkeng2021} to average the data with the same parameters as described above. The combined size of the resulting averaged CTDS backed measurement sets is apprximately 620GB, a factor of more than 6 times larger than the averaged Stokes visibilities produced by the \init\ application. The discrepancy mostly stems from the fact that the Stokes $I$ visibilities do not have a correlation axis and are compressed by \zarr\ on disk. Since the \init\ application uses the same averaging routines as those in \xova , the two averaging procedures should, however, be identical. We tested this by comparing the dirty images produced by imaging the averaged corrected data produced by \xova\ to those produced when applying the gains and averaging the data on the fly inside the \init\ application. The L2-error between the images was well within the gridding precision, in this case $10^{-7}$, confirming that the two approaches are consistent.

\subsection{Comparison between \pfb\ and \wsclean{}}
\label{sec:pfbvswsc}
\begin{figure*}
    \centering
    \includegraphics[width=2\columnwidth]{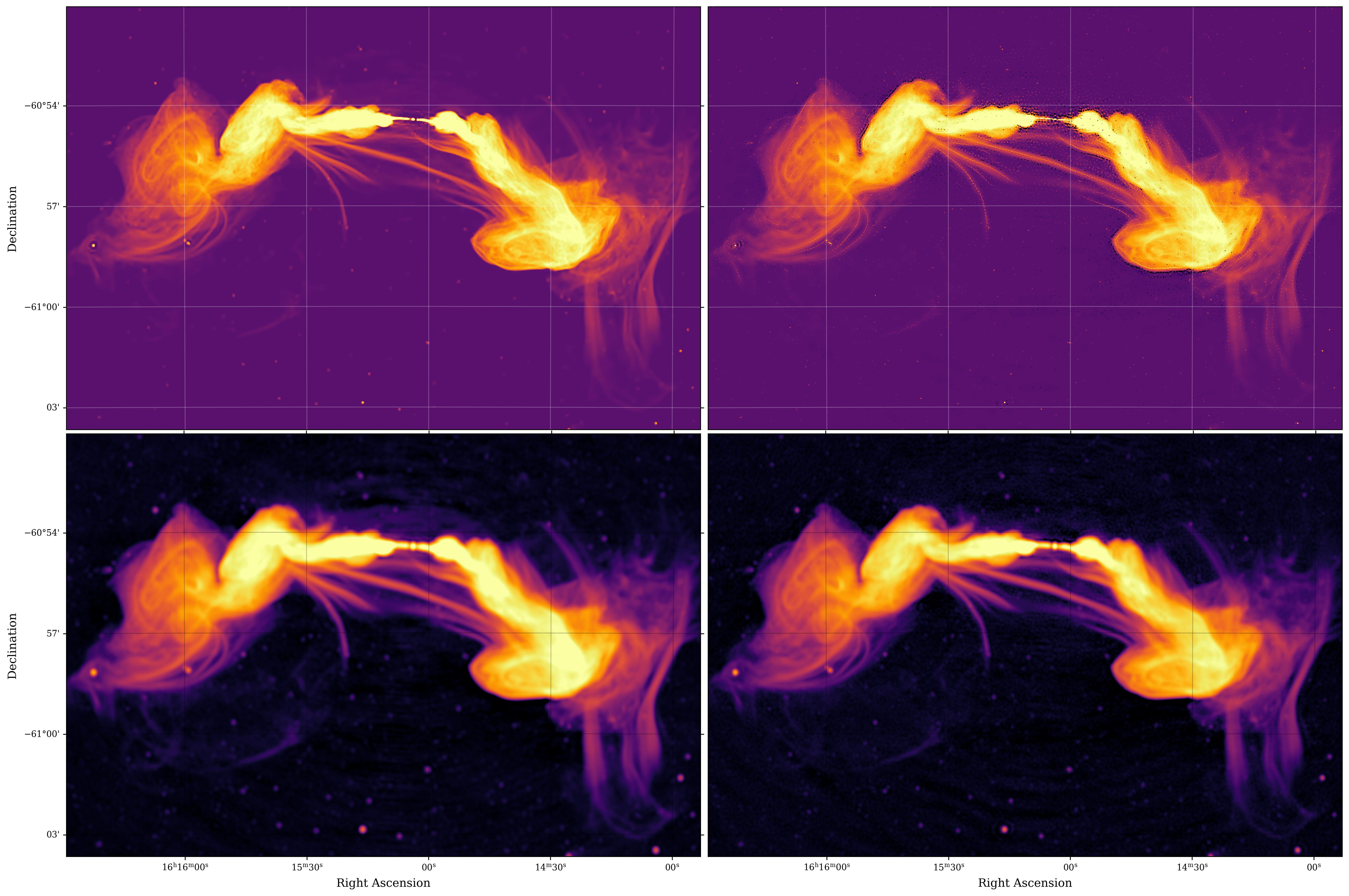}
    \caption{Comparison between \pfb\ (left) and \wsclean\ (right) of the main central source ESO137-006. Top panel shows scale matched model images in log scale. The bottom panel shows scale matched restored images also in log scale. Images are highly saturated to highlight differences.}
    \label{fig:eso006}
\end{figure*}

\begin{figure*}
    \centering
    \includegraphics[width=2\columnwidth]{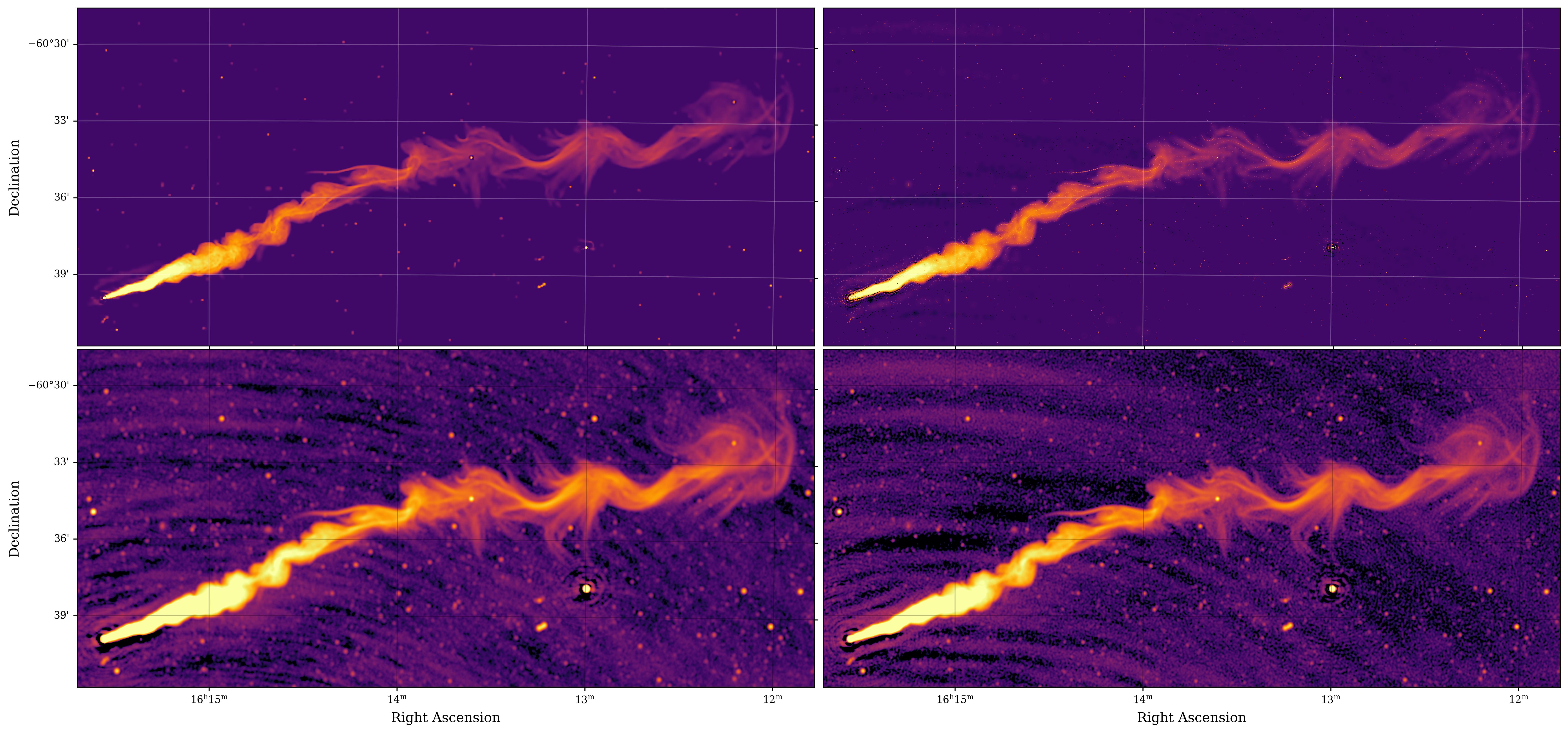}
    \caption{Comparison between \pfb\ (left) and \wsclean\ (right) for the source ESO137-007. The top panel shows scale matched model images in log scale. The bottom panel shows scaled matched restored images also in log scale. Images are highly saturated to highlight differences.}
    \label{fig:eso007}
\end{figure*}

In this section we briefly compare the performance of \pfb\ against that of \wsclean\ \cite{offringa2017multiscale}. Although an attempt was made to make the comparison as fair as possible, there are some inevitable discrepancies. For instance, even though we used the same Briggs robustness value of -0.3 for both imagers, by default \wsclean\ employs a form of multi-frequency weighting when performing joint-channel cleaning to ensure that MFS images honour the requested weighting scheme. The \pfb\ residuals are therefore slightly more naturally weighted than those from \wsclean\ but we left this parameter intact as it improved the performance of the multi-scale \clean\ algorithm. Its performance improved further by limiting the number of scales used during deconvolution to six. The parallel deconvolution option was not invoked because it introduced unphysical edges in the image. We used the auto-masking and auto-thresholding features with their parameters set to 3.5 and 1, respectively. The exact \wsclean\ command used to produce the results shown in Figures~\ref{fig:pfbvwsc}-\ref{fig:eso007} is given in $\S$~\ref{sec:wscommand}.

For this specific comparison, the \pfb\ applications were all run with their default settings which are tuned to achieve a good balance between performance and reconstruction quality for reasonably well calibrated MeerKAT data. The \stimela\ recipe used to produce these results is included in \ref{sec:recipes} and can be consulted for the exact parameter choices. Note that the default value of $\rho_{\textrm{rms}} = 1$ was used since this should, from our discussion in $\S$~\ref{sec:deconv}, preclude structures that are below the noise floor of the observation from entering the model. 

The comparison between model and residual images are shown in Figure~\ref{fig:pfbvwsc}. The top and bottom panels are scale matched and show model images in log scale and residual images in linear scale respectively. Results from \pfb\ are displayed on the left while those from \wsclean\ are shown on the right. A zoomed cutout of the central source ESO137-006 is shown in Figure~\ref{fig:eso006} with model images in the top panel and restored images on the bottom. A similar comparison is shown for the source ESO137-007 in Figure~\ref{fig:eso007}. Crucially, the differences in the restored images are negligible while the model images are completely different.

The key difference between the two approaches lies in the form of regularisation employed. The sparsity-based wavelet prior of \sara\ is better able to faithfully represent diffuse emission but it does not represent unresolved sources as individual pixels. These are represented as small Gaussian-like blobs. On the other hand, multi-scale \clean\ is able to capture unresolved sources as individual pixels, at least when these sources are located close to pixel centers, but it does not represent diffuse emission faithfully. In particular, the morphology of ESO137-006 around the central jet and the edges of the lobes is unconvincing with clear negative ridges visible in the model produced by multi-scale \clean. While such ridges could be the result of calibration artefacts and/or too coarsely discretising the measurement operator \eqref{immeasop}, it is more likely that they are artefacts introduced because of the manner in which \clean\ operates. Similar negative ridges are visible around the jet of ESO137-007 shown in Figure~\ref{fig:eso007}. However, the negative ring-like structures shown just to the north and south of the jet don't seem to be associated with real source structure but are artefacts stemming from unmodelled systematics in the data. In both cases the negative components are co-located with negative holes in the \sara\ residuals shown in the bottom left panel of Figure~\ref{fig:pfbvwsc}. This suggests that the negative holes are largely a consequence of the positivity constraint. While the constraint can be relaxed to produce comparable residuals to multi-scale \clean, doing so would defeat the purpose of trying to improve on \clean\ in the first place. Large residuals often indicate the presence of unmodelled systematics and allowing for negative ``flux" simply enables the algorithm to absorb these systematics by incorporating unphysical features into the model, which then has to be restored to get a more faithful representation of source morphology. This is counter-productive since the restoring process degrades resolution and results in a quantity that no longer fits the data. On the other hand, while the residuals corresponding to the \sara\ model have more structure to them, the morphology of the source is much more faithfully represented and with much better resolution. The deconvolution depth is, unfortunately, still limited by the presence of unmodelled systematics.

These experiments used a single node containing an AMD EPYC 7773X 64-Core Processor, 1TB of RAM and an SSD-backed file system. Table~\ref{tab:resource-usage} shows a comparison between peak and average CPU and memory utilisation as well as the runtimes for both applications\footnote{Since the node is not guaranteed to be uncontested Table~\ref{tab:resource-usage} reports the best figures obtained from three consecutive runs.}. The total runtime of \pfb\ is about double that of \wsclean{}. \pfb\ requires far more memory but utilises the CPU slightly more efficiently. Note that the metrics reported for \pfb\ include the time it takes to create the averaged Stokes visibility products whereas \wsclean\ read the corrected and averaged data directly. Approximately 80\% of the total runtime for \pfb\ is spent in the \sara\ application with FFTs and wavelet transforms accounting for the majority of floating point operations. Since both FFTs and wavelet transforms are highly amenable to hardware (e.g. GPU) acceleration, it should be possible to significantly reduce the total runtime of \sara\ by exploiting modern computing architectures. Note that the gridding costs for the two applications are about the same since they utilise the same measurement operator and the same number of major iterations (i.e. 10).
\begin{table}[htbp]
    \centering
    \small
    \begin{tabular}{@{}lccccc@{}}
    \toprule
    \multirow{2}{*}{\textbf{App}} & \multicolumn{2}{c}{\textbf{CPU (\%)}} & \multicolumn{2}{c}{\textbf{Memory (GB)}} & \multirow{2}{*}{\textbf{Runtime (hrs)}} \\
    \cmidrule(lr){2-3} \cmidrule(lr){4-5}
    & Peak & Avg & Peak & Avg & \\
    \midrule
    \pfb\ & 100 & 39 & 312 & 120 & 27 \\
    \wsclean\ & 100 & 31 & 68 & 50 & 13 \\
    \bottomrule
    \end{tabular}
    \caption{Resource utilisation for \pfb\ and \wsclean.}
    \label{tab:resource-usage}
\end{table}

\subsection{A reproducible imaging workflow}
This section illustrates how \stimela\ can be used to construct flexible and reproducible workflows for \pfb. The functionality we illustrate should be useful in situations where a suboptimal result, obtained using a standardised pipeline for example, needs to be improved to meet scientific goals. Doing so usually requires a better understanding of the unmodelled systematics affecting the data.

The main artefacts noticeable in the residuals are the ring-like feature associated with the central source (ESO137-006) and the deep negative holes around certain sources. Closer inspection of the residual image cube revealed that there are marked differences between how well the data are fit between imaging bands, with the worse residuals corresponding to bands that are more affected by RFI. We also found that up-sampling the model in frequency and computing the residuals with a higher frequency resolution eliminates the ring-like feature. This suggests that some of the remaining systematics could be due to the presence of unflagged RFI and the discretisation of the measurement operator \eqref{immeasop} along the frequency axis. To test this hypothesis, we selected out the part of the band between 1.295 GHz and 1.503 GHz that is largely free from RFI and imaged it with double the frequency resolution (i.e. 96 channels per band). We also increased the robust weighting factor from -0.3 to 0.5 to suppress short baselines less. Note that the same \stimela\ recipe could be reused simply by overwriting the default parameters from the command line. We deployed the workflow on \aws\ by augmenting the command line with the additional configuration file that is presented in Appendix B of \cite{africanus4}, to which a detailed explanation of how this mechanism works is deferred. Some of the key points to take note of at this stage are the following:
\begin{enumerate}
    \item The \init\ application was used to transfer the selected subset of the averaged Stokes visibilities to an \aws\ S3 bucket. This avoids the need to copy the raw data which is nearly two orders of magnitude larger than the subset we are considering after averaging. Note that, since the \init\ application is reading from the local file system, the raw data does not need to be compatible with object storage.
    \item Only the \grid\ and \sara\ applications (i.e. the most computationally expensive steps in the imaging workflow) are actually executed on \aws. \stimela\ takes care of launching computing resources for each application in a way that maximises efficiency and therefore reduces the overall cost. In particular, a \dask\ cluster with one c6in.8xlarge instance per imaging band was created for the \grid\ application. The \sara\ application was executed on a single c6in.24xlarge instance.
    \item The deconvolution does not start from scratch. Instead, the component model corresponding to the model shown in Figure~\ref{fig:pfbvwsc} was transferred to S3, interpolated as described in $\S$~\ref{sec:degrid}, and used to initialise the model. For this particular field, the resulting component model ends up being less than 1 GB in size. This is the only data product that needs to be retrieved from S3 after deconvolution.
\end{enumerate}
The above functionality can be useful in a number of ways. Firstly, it enables dedicated, observatory specific data centers to offload compute-intensive tasks to commodity computing platforms which are typically equipped with the most up to date hardware. This means that an observatory can plan its data center infrastructure with functionality to service the majority of expected scientific proposals without budgeting for rare and possibly unforeseeable science targets that require additional computing resources. Secondly, it enables developers to test new imaging algorithms without having to copy the full raw data while preserving a perfect record of the transformations the data have undergone thus far. The data selection mechanism also makes it possible to do this on subsets of the data. Finally, by publishing results with an accompanying versioned \stimela\ recipe that is guaranteed to run on a publicly accessible computing platform, it also aids reproducibility. Some of these topics are discussed further in \cite{africanus4}.

\begin{figure*}
    \centering
    \includegraphics[width=2\columnwidth]{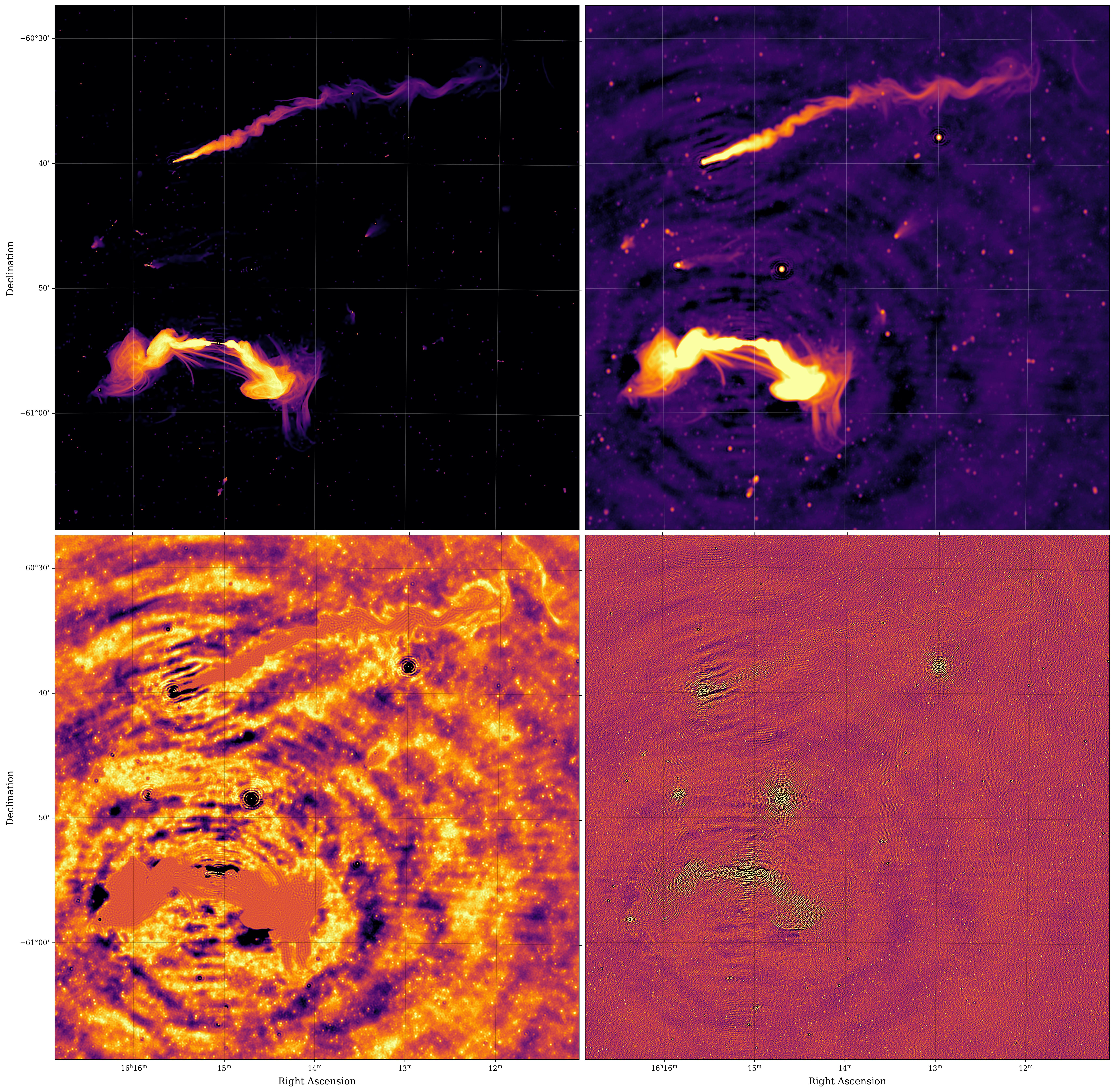}
    \caption{\pfb\ results produced on \aws\ on a subset of the ESO137 field. The top panel shows the model image on the left and the restored image on the right, both in log scale. Bottom panel shows the residual image on the left and the approximate sampling density corrected gradient \eqref{natgrad} on the right, both in linear scale. None of the images are scale matched.}
    \label{fig:pfbf}
\end{figure*}

With the component model retrieved from S3, we reran the workflow skipping the deconvolution step to produce the figures presented in Figure~\ref{fig:pfbf}. We also ran the full workflow locally on the same node that was used in the previous section. The resulting component model was compatible with the one retrieved from S3 to well within the convergence criteria of the \sara\ algorithm, verifying that the implementation on \aws\ produces the expected result. The results show that our attempt to account for the remaining unmodelled systematics was only partially successful. The main difference is that radial ripples emanating from ESO137-006 are substantially reduced by refining the frequency resolution. While the model image shown in the top left of Figure~\ref{fig:pfbf} is not really an improvement over the previous one, it should be noticeable that the resolution is not significantly affected when changing the robustness from -0.3 to 0.5. The same can't be said for the restored image shown in the top right of the figure. This is due to the scale bias introduced by weighting the data more naturally. This effect is also visible in the residuals. A comparison of the \pfb\ residuals shown in the bottom left of Figures~\ref{fig:pfbvwsc} and \ref{fig:pfbf} might actually lead one to conclude that the latter workflow performed worse. This is, however, just a trick of the weighting scheme. This cautions against using residual images as the main criteria by which to judge the quality of imaging results. The approximate sampling density corrected gradient, i.e. $U_Z^{-1} \nabla f(x_k)$ at the final iteration, shown in the bottom right of Figure \ref{fig:pfbf}, is a slightly better metric since it does not suffer from the same scale bias. This image still shows clear artefacts which limits deconvolution depth but this is only partially due to remaining issues with the data. Given the ill-posed nature of the problem, prior specification also has a big role to play. The regularisation utilised for the current work is more suitable for extended emission and does not perform as well on unresolved sources. In particular, bright point sources, especially when embedded in extended emission, tend to have empty bowls around them. Many of the more recent algorithms that have been proposed in the literature suffer from the same limitation. In fact, we are not aware of any algorithm that can simultaneously and automatically reconstruct both point sources and extended emission faithfully, emphasising the need for a framework that enables rapid prototyping and development.

\subsection{Generalisation}
\begin{figure*}
    \centering
    \includegraphics[width=2\columnwidth]{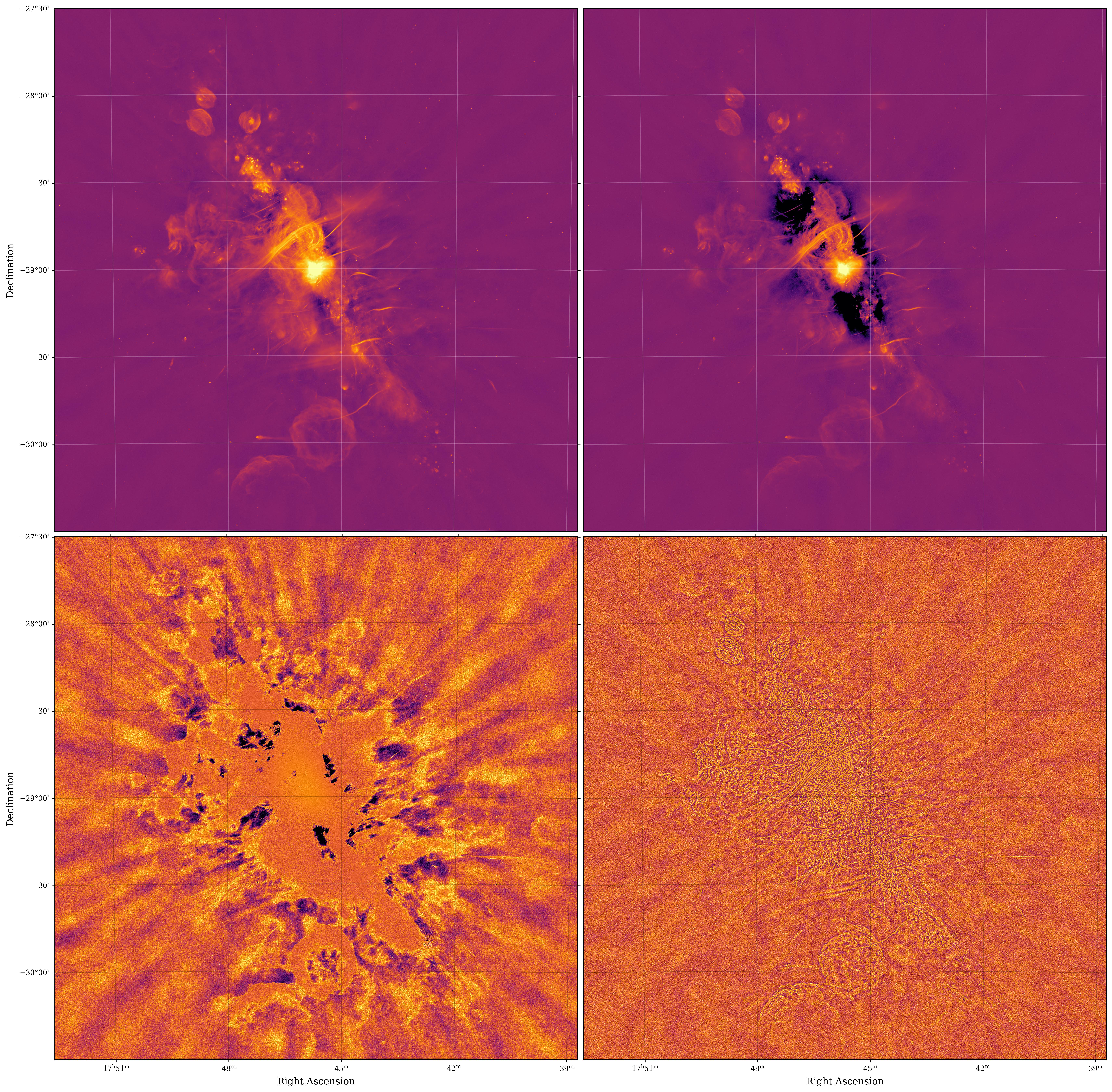}
    \caption{Comparison between \pfb\ (left) and \wsclean\ (right) on the SGRA field before commencing self-calibration. Top panel shows scale matched restored images in log scale. The bottom panel shows scale matched residual images in linear scale.}
    \label{fig:pfbvwsc_sgra}
\end{figure*}

As a final test for our imaging framework, we investigate how the imaging strategies encoded in the \stimela\ recipes given in \ref{sec:recipes} generalise to a different observation. We selected an observation of the Sagittarius A (SGRA) complex in the galactic center taken with MeerKAT at UHF band (0.544 GHz-1.087 GHz) (project ID SSV-20200519-FC-01) for this test. The observation was pre-processed using \textsc{oxkat}\footnote{\url{https://github.com/IanHeywood/oxkat}}. The corrected data after initial transfer calibration was imaged with \wsclean\ and \pfb\ using the same settings as those used in $\S$~\ref{sec:pfbvswsc}, except with imaging parameters more suitable for UHF band (i.e. $1.7^{\prime\prime} \times 1.7^{\prime\prime}$ pixels and a bandwidth of 36.7 MHz per imaging band) and a robustness value of $-1$. This resulted in a $12 \times 7500 \times 7500$ image covering approximately $3.5^\circ$ field of view over the entire unflagged part of the band. The resulting residual and restored images are displayed in Figure~\ref{fig:pfbvwsc_sgra}, again with \pfb\ results shown on the left and \wsclean\ results displayed on the right.
The deep negative bowls around the galactic bulge in the multi-scale \clean\ result suggest that the \sara\ algorithm outperforms multi-scale \clean\ on this observation. While it is possible to improve on the result from \wsclean{}, doing so requires extensive manual tuning (e.g. iteratively constructing masks and manually selecting appropriate scales to use for the multi-scale algorithm) which requires expertise on the part of the user. This is not to say that the \sara\ algorithm always outperforms multi-scale \clean{}. Prior specification is by its very nature subjective and different priors will be suitable for different kinds of observations. Because of its speed and simplicity, \clean\ remains the algorithm of choice for sparsely populated fields that are dominated by compact emission. It is not, however, suitable for fields dominated by extended emission with complex morphologies. Such fields require more sophisticated, and often more computationally expensive, algorithms to get a reliable result, especially if an automated pipeline is used for the initial processing. The \sara\ algorithm detailed in $\S$~\ref{sec:deconv} is but one example of such an algorithm. The fact that it enjoys convergence guarantees, has interpretable hyper-parameters and does not require manually constructing masks, makes it particularly attractive for use in automated imaging pipelines. While the algorithm is more computationally expensive than \clean{}, we have illustrated that, perhaps contrary to popular belief, it does not require a high performance computing cluster for MeerKAT sized problems.

\section{Conclusion}
\label{sec:conclusion}
This paper presents a general imaging framework that can be used for the interferometric imaging problem in the presence of known direction-independent antenna gains. We illustrated an application of the proposed framework by processing TB scale data from the MeerKAT relying solely on tools that are available in the \africanus\ ecosystem. In so doing, we touched on a number of problems that are set to become more severe as modern radio interferometric arrays push us ever further into the big data regime.

First and foremost, the imaging framework we present aims for maximum flexibility and ease of development while keeping computational costs reasonable. We believe this to be an important consideration given the ambitious and forever evolving scientific goals of upcoming projects. The reason for this is that the algorithms that are typically employed to process radio interferometric data are, largely because they are often tuned for performance rather than accuracy, suboptimal. Imaging, in particular, is not a solved problem. More sophisticated algorithms enable better science which make them interesting not only for upcoming projects but also for revisiting existing archives. At the same time, data sizes, bespoke data formats and the complexity of the radio interferometric imaging problem sets the barrier to entry for algorithm developers astoundingly high. \pfb\ attempts to lower this barrier to facilitate accelerating algorithm development. The increased flexibility comes at the expense of speed and efficiency but, as illustrated in this paper, it is still possible to develop practical solutions within this framework. Once a useful algorithm has been identified it can, and should, be implemented in a more efficient low-level programming language. Indeed, part of the motivation for deconstructing \pfb\ into separate applications is to make it easier to do this incrementally.

We also touched on the need for intermediary data products during imaging. As shown in $\S$~\ref{sec:measmod}, when considering the imaging and calibration problems separately, the computational complexity of the imaging problem can be significantly reduced by transforming the data into a form that obviates the need to apply the gains at every iteration and which is more amenable to averaging. This has to be contrasted to the more typical approach of forming corrected data products. While the two procedures yield identical results, our approach is grounded in statistical reasoning instead of inverse modelling and therefore lends itself more readily to interpretation. Firstly, it should be clear that this is a computational shortcut that is necessitated by the size of the data involved. Since discarding the off-diagonal elements of the Mueller weights presupposes that the Stokes\footnote{This argument can also be phrased in terms of the coherencies associated with elements of the brightness matrix $\bm{X}_{pq}$ which would be better for joint polarisation imaging but such is out of scope for the current discussion.} parameters are independent, it is by no means optimal. By applying the gains and subsequently performing averaging on the fly using the {\tt init} application, we simultaneously avoid the need to duplicate the data at the original resolution and maintain a perfect record of the transformations that the data undergo during the imaging procedure. This makes the intermediary data products truly disposable since the only additional data products that need to be stored to reproduce the experiment are the gain tables and the model datasets, which are usually much smaller than the raw data. It also facilitates simultaneously experimenting with different calibration and/or imaging models since the original data only need to be read at the outset. In combination with the data selection mechanisms discussed in $\S$~\ref{sec:init}, this should enable rapid prototyping of new algorithms without the need for unnecessarily large intermediary data products.

Even though we focused on a particular sparsity-based imaging algorithm, the preconditioned forward-backward algorithmic framework we utilise is by no means limited to sparsity-based imaging. In the context of interferometric imaging algorithms, it allows generalising the concept of a minor cycle to almost arbitrary regularisers. By virtue of being rooted in forward modelling, it also makes it possible to account for certain systematics in a more robust way than is possible with \clean. For example, since the operators defined in $\S$~\ref{sec:measmod} can be partitioned in time, the time variability of the primary beam pattern can be accounted for to some extent. Partitioning by time would also improve the accuracy of the PSF approximation \eqref{hessapprox} since the spread in the $w$ coordinate typically increases with observing time. If combined with a rephasing operation that aligns multiple pointings to the same phase center, and suitable primary beam interpolation, this same mechanism could be used to perform on the fly mosaicing in a way that is far more accurate than the traditional approach of imaging each field separately and stitching primary beam corrected images together after the fact. This is an avenue we intend to pursue in future work.

Of course, encoding additional systematics into the measurement operator comes with increased complexity and associated computational costs. This is one of the reasons for considering commodity computing platforms such as \aws. The majority of science targets likely won't require the latest and greatest algorithmic and technological advances but, since radio interferometers have a way of turning up surprises, there will always be high profile science that is beyond the reach of standardised workflows. The ability to process observations with a highly customisable software suite without getting bogged down by limited computing infrastructure could be extremely valuable in such situations. It is with this outlook that \pfb, and the \africanus\ ecosystem as a whole, is being developed.

\section*{Acknowledgements}
We would like to thank Philipp Arras, Philipp Frank and Ming Jiang for useful discussions.
The MeerKAT telescope is operated by the South African Radio Astronomy Observatory, which is a facility of the National Research Foundation, an agency of the Department of Science and Innovation.
Funding: OMS's and JSK's research is supported by the South African Research Chairs Initiative of the Department of Science and Technology and National Research Foundation (grant No. 81737). YW was supported by the UK Research and Innovation under the EPSRC grant EP/T028270/1 and the STFC grant ST/W000970/1. JR acknowledges financial support from the German Federal Ministry of Education and Research (BMBF) under grant 05A23WO1 (Verbundprojekt
D-MeerKAT III).

\paragraph*{Data Availability} The MeerKAT data used for the benchmarks in this paper series is publicly available via the SARAO archive\footnote{\url{https://archive.sarao.ac.za}}. Specifically, the ESO137 and SGRA data are available through the project IDs SCI-20190418-SM-01 and SSV-20200519-FC-01, respectively.

\bibliographystyle{elsarticle-harv}
\bibliography{pfb-imaging.bib}

\begin{thebibliography}{60}
\expandafter\ifx\csname natexlab\endcsname\relax\def\natexlab#1{#1}\fi
\providecommand{\url}[1]{\texttt{#1}}
\providecommand{\href}[2]{#2}
\providecommand{\path}[1]{#1}
\providecommand{\DOIprefix}{doi:}
\providecommand{\ArXivprefix}{arXiv:}
\providecommand{\URLprefix}{URL: }
\providecommand{\Pubmedprefix}{pmid:}
\providecommand{\doi}[1]{\href{http://dx.doi.org/#1}{\path{#1}}}
\providecommand{\Pubmed}[1]{\href{pmid:#1}{\path{#1}}}
\providecommand{\bibinfo}[2]{#2}
\ifx\xfnm\relax \def\xfnm[#1]{\unskip,\space#1}\fi
%Type = Article
\bibitem[{{Abdulaziz} et~al.(2019){Abdulaziz}, {Dabbech} and
  {Wiaux}}]{hypersara}
\bibinfo{author}{{Abdulaziz}, A.}, \bibinfo{author}{{Dabbech}, A.},
  \bibinfo{author}{{Wiaux}, Y.}, \bibinfo{year}{2019}.
\newblock \bibinfo{title}{{Wideband super-resolution imaging in Radio
  Interferometry via low rankness and joint average sparsity models
  (HyperSARA)}}.
\newblock \bibinfo{journal}{MNRAS} \bibinfo{volume}{489},
  \bibinfo{pages}{1230--1248}.
\newblock \DOIprefix\doi{10.1093/mnras/stz2117},
  \href{http://arxiv.org/abs/1806.04596}{{\tt arXiv:1806.04596}}.
%Type = Inproceedings
\bibitem[{{Abernathey} et~al.(2018){Abernathey}, {Hamman} and
  {Miles}}]{Zarr2018}
\bibinfo{author}{{Abernathey}, R.P.}, \bibinfo{author}{{Hamman}, J.},
  \bibinfo{author}{{Miles}, A.}, \bibinfo{year}{2018}.
\newblock \bibinfo{title}{{Beyond netCDF: Cloud Native Climate Data with Zarr
  and XArray}}, in: \bibinfo{booktitle}{AGU Fall Meeting Abstracts}, pp.
  \bibinfo{pages}{IN33A--06}.
%Type = Manual
\bibitem[{{Apache Software Foundation.}(2019)}]{Arrow2019}
\bibinfo{author}{{Apache Software Foundation.}}, \bibinfo{year}{2019}.
\newblock \bibinfo{title}{Arrow: a cross-language development platform for
  in-memory data}.
%Type = Article
\bibitem[{{Arras} et~al.(2021){Arras}, {Bester}, {Perley}, {Leike}, {Smirnov},
  {Westermann} and {En{\ss}lin}}]{Arras_2021}
\bibinfo{author}{{Arras}, P.}, \bibinfo{author}{{Bester}, H.L.},
  \bibinfo{author}{{Perley}, R.A.}, \bibinfo{author}{{Leike}, R.},
  \bibinfo{author}{{Smirnov}, O.}, \bibinfo{author}{{Westermann}, R.},
  \bibinfo{author}{{En{\ss}lin}, T.A.}, \bibinfo{year}{2021}.
\newblock \bibinfo{title}{{Comparison of classical and Bayesian imaging in
  radio interferometry. Cygnus A with CLEAN and resolve}}.
\newblock \bibinfo{journal}{A\& A} \bibinfo{volume}{646}, \bibinfo{pages}{A84}.
\newblock \DOIprefix\doi{10.1051/0004-6361/202039258}.
%Type = Article
\bibitem[{Arras et~al.(2021)Arras, Reinecke, Westermann and
  En{\ss}in}]{wgridder}
\bibinfo{author}{Arras, P.}, \bibinfo{author}{Reinecke, M.},
  \bibinfo{author}{Westermann, R.}, \bibinfo{author}{En{\ss}in, T.A.},
  \bibinfo{year}{2021}.
\newblock \bibinfo{title}{Efficient wide-field radio interferometry response}.
\newblock \bibinfo{journal}{A\& A} \bibinfo{volume}{646}, \bibinfo{pages}{A58}.
\newblock \DOIprefix\doi{10.1051/0004-6361/202039723}.
%Type = Misc
\bibitem[{Atemkeng et~al.(2021)Atemkeng, Perkins, Kenyon, Hugo and
  Smirnov}]{atemkeng2021}
\bibinfo{author}{Atemkeng, M.}, \bibinfo{author}{Perkins, S.},
  \bibinfo{author}{Kenyon, J.}, \bibinfo{author}{Hugo, B.},
  \bibinfo{author}{Smirnov, O.}, \bibinfo{year}{2021}.
\newblock \bibinfo{title}{Xova: Baseline-dependent time and channel averaging
  for radio interferometry}.
\newblock \href{http://arxiv.org/abs/2101.11270}{{\tt arXiv:2101.11270}}.
%Type = Article
\bibitem[{{Barnett} et~al.(2018){Barnett}, {Magland} and
  {Klinteberg}}]{barnett_kernel}
\bibinfo{author}{{Barnett}, A.H.}, \bibinfo{author}{{Magland}, J.F.},
  \bibinfo{author}{{Klinteberg}, L.a.}, \bibinfo{year}{2018}.
\newblock \bibinfo{title}{{A parallel non-uniform fast Fourier transform
  library based on an ``exponential of semicircle'' kernel}}.
\newblock \bibinfo{journal}{arXiv e-prints} ,
  \bibinfo{pages}{arXiv:1808.06736}\href{http://arxiv.org/abs/1808.06736}{{\tt
  arXiv:1808.06736}}.
%Type = Article
\bibitem[{{Bester} et~al.(2021){Bester}, {Repetti}, {Perkins}, {Smirnov} and
  {Kenyon}}]{bester2021}
\bibinfo{author}{{Bester}, H.L.}, \bibinfo{author}{{Repetti}, A.},
  \bibinfo{author}{{Perkins}, S.}, \bibinfo{author}{{Smirnov}, O.M.},
  \bibinfo{author}{{Kenyon}, J.S.}, \bibinfo{year}{2021}.
\newblock \bibinfo{title}{{A practical preconditioner for wide-field continuum
  imaging of radio interferometric data}}.
\newblock \bibinfo{journal}{arXiv e-prints} ,
  \bibinfo{pages}{arXiv:2101.08072}\DOIprefix\doi{10.48550/arXiv.2101.08072},
  \href{http://arxiv.org/abs/2101.08072}{{\tt arXiv:2101.08072}}.
%Type = Inproceedings
\bibitem[{Brewer(2015)}]{brewer2015}
\bibinfo{author}{Brewer, E.A.}, \bibinfo{year}{2015}.
\newblock \bibinfo{title}{Kubernetes and the path to cloud native}, in:
  \bibinfo{booktitle}{Proceedings of the Sixth ACM Symposium on Cloud
  Computing}, \bibinfo{publisher}{Association for Computing Machinery},
  \bibinfo{address}{New York, NY, USA}. p. \bibinfo{pages}{167}.
\newblock \DOIprefix\doi{10.1145/2806777.2809955}.
%Type = Phdthesis
\bibitem[{Briggs(1995)}]{briggs1995high}
\bibinfo{author}{Briggs, D.S.}, \bibinfo{year}{1995}.
\newblock \bibinfo{title}{High Fidelity Deconvolution of Moderately Resolved
  Sources}.
\newblock Ph.D. thesis. New Mexico Institute of Mining and Technology.
  \bibinfo{address}{Socorro, New Mexico}.
\newblock \bibinfo{note}{Ph.D. Dissertation}.
%Type = Article
\bibitem[{{Cai} et~al.(2018){Cai}, {Pereyra} and {McEwen}}]{Cai2018I}
\bibinfo{author}{{Cai}, X.}, \bibinfo{author}{{Pereyra}, M.},
  \bibinfo{author}{{McEwen}, J.D.}, \bibinfo{year}{2018}.
\newblock \bibinfo{title}{{Uncertainty quantification for radio interferometric
  imaging - I. Proximal MCMC methods}}.
\newblock \bibinfo{journal}{MNRAS} \bibinfo{volume}{480},
  \bibinfo{pages}{4154--4169}.
\newblock \DOIprefix\doi{10.1093/mnras/sty2004},
  \href{http://arxiv.org/abs/1711.04818}{{\tt arXiv:1711.04818}}.
%Type = Article
\bibitem[{{Candes} et~al.(2007){Candes}, {Wakin} and {Boyd}}]{candes07}
\bibinfo{author}{{Candes}, E.J.}, \bibinfo{author}{{Wakin}, M.B.},
  \bibinfo{author}{{Boyd}, S.P.}, \bibinfo{year}{2007}.
\newblock \bibinfo{title}{{Enhancing Sparsity by Reweighted L1 Minimization}}.
\newblock \bibinfo{journal}{arXiv e-prints} ,
  \bibinfo{pages}{arXiv:0711.1612}\href{http://arxiv.org/abs/0711.1612}{{\tt
  arXiv:0711.1612}}.
%Type = Article
\bibitem[{{Carrillo} et~al.(2012){Carrillo}, {McEwen} and {Wiaux}}]{sara}
\bibinfo{author}{{Carrillo}, R.E.}, \bibinfo{author}{{McEwen}, J.D.},
  \bibinfo{author}{{Wiaux}, Y.}, \bibinfo{year}{2012}.
\newblock \bibinfo{title}{{Sparsity Averaging Reweighted Analysis (SARA): a
  novel algorithm for radio-interferometric imaging}}.
\newblock \bibinfo{journal}{MNRAS} \bibinfo{volume}{426},
  \bibinfo{pages}{1223--1234}.
\newblock \DOIprefix\doi{10.1111/j.1365-2966.2012.21605.x},
  \href{http://arxiv.org/abs/1205.3123}{{\tt arXiv:1205.3123}}.
%Type = Article
\bibitem[{Condat(2013)}]{condat2012}
\bibinfo{author}{Condat, L.}, \bibinfo{year}{2013}.
\newblock \bibinfo{title}{{A primal-dual splitting method for convex
  optimization involving Lipschitzian, proximable and linear composite terms}}.
\newblock \bibinfo{journal}{{Journal of Optimization Theory and Applications}}
  , \bibinfo{pages}{online first, to
  appear}\DOIprefix\doi{10.1007/s10957-012-0245-9}.
%Type = Article
\bibitem[{Cornwell and Perley(1992)}]{cornwell1992radio}
\bibinfo{author}{Cornwell, T.}, \bibinfo{author}{Perley, R.},
  \bibinfo{year}{1992}.
\newblock \bibinfo{title}{Radio-interferometric imaging of very large
  fields-the problem of non-coplanar arrays}.
\newblock \bibinfo{journal}{Astronomy and Astrophysics} \bibinfo{volume}{261},
  \bibinfo{pages}{353--364}.
%Type = Article
\bibitem[{Cornwell and Wilkinson(1981)}]{cornwell1981new}
\bibinfo{author}{Cornwell, T.}, \bibinfo{author}{Wilkinson, P.},
  \bibinfo{year}{1981}.
\newblock \bibinfo{title}{A new method for making maps with unstable radio
  interferometers}.
\newblock \bibinfo{journal}{Monthly Notices of the Royal Astronomical Society}
  \bibinfo{volume}{196}, \bibinfo{pages}{1067--1086}.
%Type = Article
\bibitem[{Daubechies(1988)}]{Daubechies1988}
\bibinfo{author}{Daubechies, I.}, \bibinfo{year}{1988}.
\newblock \bibinfo{title}{Orthonormal bases of compactly supported wavelets}.
\newblock \bibinfo{journal}{Communications on Pure and Applied Mathematics}
  \bibinfo{volume}{41}, \bibinfo{pages}{909--996}.
\newblock \DOIprefix\doi{10.1002/cpa.3160410705}.
%Type = Inproceedings
\bibitem[{{Di Francesco} et~al.(2019){Di Francesco}, {Chalmers}, {Denman},
  {Fissel}, {Friesen}, {Gaensler}, {Hlavacek-Larrondo}, {Kirk}, {Matthews},
  {O'Dea}, {Robishaw}, {Rosolowsky}, {Rupen}, {Sadavoy}, {Sa-Harb}, {Sivakoff},
  {Tahani}, {van der Marel}, {White} and {Wilson}}]{ngvla}
\bibinfo{author}{{Di Francesco}, J.}, \bibinfo{author}{{Chalmers}, D.},
  \bibinfo{author}{{Denman}, N.}, \bibinfo{author}{{Fissel}, L.},
  \bibinfo{author}{{Friesen}, R.}, \bibinfo{author}{{Gaensler}, B.},
  \bibinfo{author}{{Hlavacek-Larrondo}, J.}, \bibinfo{author}{{Kirk}, H.},
  \bibinfo{author}{{Matthews}, B.}, \bibinfo{author}{{O'Dea}, C.},
  \bibinfo{author}{{Robishaw}, T.}, \bibinfo{author}{{Rosolowsky}, E.},
  \bibinfo{author}{{Rupen}, M.}, \bibinfo{author}{{Sadavoy}, S.},
  \bibinfo{author}{{Sa-Harb}, S.}, \bibinfo{author}{{Sivakoff}, G.},
  \bibinfo{author}{{Tahani}, M.}, \bibinfo{author}{{van der Marel}, N.},
  \bibinfo{author}{{White}, J.}, \bibinfo{author}{{Wilson}, C.},
  \bibinfo{year}{2019}.
\newblock \bibinfo{title}{{The Next Generation Very Large Array}}, in:
  \bibinfo{booktitle}{Canadian Long Range Plan for Astronomy and Astrophysics
  White Papers}, p.~\bibinfo{pages}{32}.
\newblock \DOIprefix\doi{10.5281/zenodo.3765763},
  \href{http://arxiv.org/abs/1911.01517}{{\tt arXiv:1911.01517}}.
%Type = Book
\bibitem[{Goldstein(2003)}]{Goldstein2003PolarizedLight}
\bibinfo{author}{Goldstein, D.H.}, \bibinfo{year}{2003}.
\newblock \bibinfo{title}{Polarized Light}.
\newblock \bibinfo{edition}{2nd} ed., \bibinfo{publisher}{Marcel Dekker},
  \bibinfo{address}{New York}.
\newblock \bibinfo{note}{Chapter 3 covers Mueller calculus in detail}.
%Type = Misc
\bibitem[{Hallinan et~al.(2019)Hallinan, Ravi, Weinreb, Kocz, Huang, Woody,
  Lamb, D'Addario, Catha, Shi, Law, Kulkarni, Phinney, Eastwood, Bouman,
  McLaughlin, Ransom, Siemens, Cordes, Lynch, Kaplan, Chatterjee, Lazio,
  Brazier, Bhatnagar, Myers, Walter and Gaensler}]{dsa2000}
\bibinfo{author}{Hallinan, G.}, \bibinfo{author}{Ravi, V.},
  \bibinfo{author}{Weinreb, S.}, \bibinfo{author}{Kocz, J.},
  \bibinfo{author}{Huang, Y.}, \bibinfo{author}{Woody, D.P.},
  \bibinfo{author}{Lamb, J.}, \bibinfo{author}{D'Addario, L.},
  \bibinfo{author}{Catha, M.}, \bibinfo{author}{Shi, J.}, \bibinfo{author}{Law,
  C.}, \bibinfo{author}{Kulkarni, S.R.}, \bibinfo{author}{Phinney, E.S.},
  \bibinfo{author}{Eastwood, M.W.}, \bibinfo{author}{Bouman, K.L.},
  \bibinfo{author}{McLaughlin, M.A.}, \bibinfo{author}{Ransom, S.M.},
  \bibinfo{author}{Siemens, X.}, \bibinfo{author}{Cordes, J.M.},
  \bibinfo{author}{Lynch, R.S.}, \bibinfo{author}{Kaplan, D.L.},
  \bibinfo{author}{Chatterjee, S.}, \bibinfo{author}{Lazio, J.},
  \bibinfo{author}{Brazier, A.}, \bibinfo{author}{Bhatnagar, S.},
  \bibinfo{author}{Myers, S.T.}, \bibinfo{author}{Walter, F.},
  \bibinfo{author}{Gaensler, B.M.}, \bibinfo{year}{2019}.
\newblock \bibinfo{title}{The dsa-2000 -- a radio survey camera}.
\newblock \DOIprefix\doi{10.48550/ARXIV.1907.07648}.
%Type = Article
\bibitem[{{Hamaker} et~al.(1996){Hamaker}, {Bregman} and
  {Sault}}]{hamaker1996understanding1}
\bibinfo{author}{{Hamaker}, J.P.}, \bibinfo{author}{{Bregman}, J.D.},
  \bibinfo{author}{{Sault}, R.J.}, \bibinfo{year}{1996}.
\newblock \bibinfo{title}{{Understanding radio polarimetry. I. Mathematical
  foundations.}}
\newblock \bibinfo{journal}{A\& As} \bibinfo{volume}{117},
  \bibinfo{pages}{137--147}.
%Type = Article
\bibitem[{Harris et~al.(2020)Harris, Millman, van~der Walt, Gommers, Virtanen,
  Cournapeau, Wieser, Taylor, Berg, Smith, Kern, Picus, Hoyer, van Kerkwijk,
  Brett, Haldane, del R{'{\i}}o, Wiebe, Peterson, G{'{e}}rard-Marchant,
  Sheppard, Reddy, Weckesser, Abbasi, Gohlke and Oliphant}]{Harris2020}
\bibinfo{author}{Harris, C.R.}, \bibinfo{author}{Millman, K.J.},
  \bibinfo{author}{van~der Walt, S.J.}, \bibinfo{author}{Gommers, R.},
  \bibinfo{author}{Virtanen, P.}, \bibinfo{author}{Cournapeau, D.},
  \bibinfo{author}{Wieser, E.}, \bibinfo{author}{Taylor, J.},
  \bibinfo{author}{Berg, S.}, \bibinfo{author}{Smith, N.J.},
  \bibinfo{author}{Kern, R.}, \bibinfo{author}{Picus, M.},
  \bibinfo{author}{Hoyer, S.}, \bibinfo{author}{van Kerkwijk, M.H.},
  \bibinfo{author}{Brett, M.}, \bibinfo{author}{Haldane, A.},
  \bibinfo{author}{del R{'{\i}}o, J.F.}, \bibinfo{author}{Wiebe, M.},
  \bibinfo{author}{Peterson, P.}, \bibinfo{author}{G{'{e}}rard-Marchant, P.},
  \bibinfo{author}{Sheppard, K.}, \bibinfo{author}{Reddy, T.},
  \bibinfo{author}{Weckesser, W.}, \bibinfo{author}{Abbasi, H.},
  \bibinfo{author}{Gohlke, C.}, \bibinfo{author}{Oliphant, T.E.},
  \bibinfo{year}{2020}.
\newblock \bibinfo{title}{Array programming with {NumPy}}.
\newblock \bibinfo{journal}{Nature} \bibinfo{volume}{585},
  \bibinfo{pages}{357--362}.
\newblock \DOIprefix\doi{10.1038/s41586-020-2649-2}.
%Type = Article
\bibitem[{{H{\"o}gbom}(1974)}]{clean1974}
\bibinfo{author}{{H{\"o}gbom}, J.A.}, \bibinfo{year}{1974}.
\newblock \bibinfo{title}{{Aperture Synthesis with a Non-Regular Distribution
  of Interferometer Baselines}}.
\newblock \bibinfo{journal}{A\& As} \bibinfo{volume}{15}, \bibinfo{pages}{417}.
%Type = Article
\bibitem[{Hotan et~al.(2021)Hotan, Bunton, Chippendale, Whiting, Tuthill, Moss,
  McConnell, Amy, Huynh, Allison, Anderson, Bannister, Bastholm, Beresford,
  Bock, Bolton, Chapman, Chow, Collier, Cooray, Cornwell, Diamond, Edwards,
  Feain, Franzen, George, Gupta, Hampson, Harvey-Smith, Hayman, Heywood, Jacka,
  Jackson, Jackson, Jeganathan, Johnston, Kesteven, Kleiner, Koribalski,
  Lee-Waddell, Lenc, Lensson, Mackay, Mahony, McClure-Griffiths, McConigley,
  Mirtschin, Ng, Norris, Pearce, Phillips, Pilawa, Raja, Reynolds, Roberts,
  Roxby, Sadler, Shields, Schinckel, Serra, Shaw, Sweetnam, Troup, Tzioumis,
  Voronkov and Westmeier}]{Hotan_2021}
\bibinfo{author}{Hotan, A.W.}, \bibinfo{author}{Bunton, J.D.},
  \bibinfo{author}{Chippendale, A.P.}, \bibinfo{author}{Whiting, M.},
  \bibinfo{author}{Tuthill, J.}, \bibinfo{author}{Moss, V.A.},
  \bibinfo{author}{McConnell, D.}, \bibinfo{author}{Amy, S.W.},
  \bibinfo{author}{Huynh, M.T.}, \bibinfo{author}{Allison, J.R.},
  \bibinfo{author}{Anderson, C.S.}, \bibinfo{author}{Bannister, K.W.},
  \bibinfo{author}{Bastholm, E.}, \bibinfo{author}{Beresford, R.},
  \bibinfo{author}{Bock, D.C.J.}, \bibinfo{author}{Bolton, R.},
  \bibinfo{author}{Chapman, J.M.}, \bibinfo{author}{Chow, K.},
  \bibinfo{author}{Collier, J.D.}, \bibinfo{author}{Cooray, F.R.},
  \bibinfo{author}{Cornwell, T.J.}, \bibinfo{author}{Diamond, P.J.},
  \bibinfo{author}{Edwards, P.G.}, \bibinfo{author}{Feain, I.J.},
  \bibinfo{author}{Franzen, T.M.O.}, \bibinfo{author}{George, D.},
  \bibinfo{author}{Gupta, N.}, \bibinfo{author}{Hampson, G.A.},
  \bibinfo{author}{Harvey-Smith, L.}, \bibinfo{author}{Hayman, D.B.},
  \bibinfo{author}{Heywood, I.}, \bibinfo{author}{Jacka, C.},
  \bibinfo{author}{Jackson, C.A.}, \bibinfo{author}{Jackson, S.},
  \bibinfo{author}{Jeganathan, K.}, \bibinfo{author}{Johnston, S.},
  \bibinfo{author}{Kesteven, M.}, \bibinfo{author}{Kleiner, D.},
  \bibinfo{author}{Koribalski, B.S.}, \bibinfo{author}{Lee-Waddell, K.},
  \bibinfo{author}{Lenc, E.}, \bibinfo{author}{Lensson, E.S.},
  \bibinfo{author}{Mackay, S.}, \bibinfo{author}{Mahony, E.K.},
  \bibinfo{author}{McClure-Griffiths, N.M.}, \bibinfo{author}{McConigley, R.},
  \bibinfo{author}{Mirtschin, P.}, \bibinfo{author}{Ng, A.K.},
  \bibinfo{author}{Norris, R.P.}, \bibinfo{author}{Pearce, S.E.},
  \bibinfo{author}{Phillips, C.}, \bibinfo{author}{Pilawa, M.A.},
  \bibinfo{author}{Raja, W.}, \bibinfo{author}{Reynolds, J.E.},
  \bibinfo{author}{Roberts, P.}, \bibinfo{author}{Roxby, D.N.},
  \bibinfo{author}{Sadler, E.M.}, \bibinfo{author}{Shields, M.},
  \bibinfo{author}{Schinckel, A.E.T.}, \bibinfo{author}{Serra, P.},
  \bibinfo{author}{Shaw, R.D.}, \bibinfo{author}{Sweetnam, T.},
  \bibinfo{author}{Troup, E.R.}, \bibinfo{author}{Tzioumis, A.},
  \bibinfo{author}{Voronkov, M.A.}, \bibinfo{author}{Westmeier, T.},
  \bibinfo{year}{2021}.
\newblock \bibinfo{title}{Australian square kilometre array pathfinder: I.
  system description}.
\newblock \bibinfo{journal}{Publications of the Astronomical Society of
  Australia} \bibinfo{volume}{38}.
\newblock \DOIprefix\doi{10.1017/pasa.2021.1}.
%Type = Article
\bibitem[{Hoyer and Hamman(2017)}]{hoyer2017}
\bibinfo{author}{Hoyer, S.}, \bibinfo{author}{Hamman, J.},
  \bibinfo{year}{2017}.
\newblock \bibinfo{title}{xarray: {N-D} labeled arrays and datasets in
  {Python}}.
\newblock \bibinfo{journal}{Journal of Open Research Software}
  \bibinfo{volume}{5}.
\newblock \DOIprefix\doi{10.5334/jors.148}.
%Type = Article
\bibitem[{Hunter and Lange(2004)}]{mmtut}
\bibinfo{author}{Hunter, D.R.}, \bibinfo{author}{Lange, K.},
  \bibinfo{year}{2004}.
\newblock \bibinfo{title}{A tutorial on mm algorithms}.
\newblock \bibinfo{journal}{The American Statistician} \bibinfo{volume}{58},
  \bibinfo{pages}{30--37}.
\newblock \DOIprefix\doi{10.1198/0003130042836},
  \href{http://arxiv.org/abs/https://doi.org/10.1198/0003130042836}{{\tt
  arXiv:https://doi.org/10.1198/0003130042836}}.
%Type = Inproceedings
\bibitem[{{Jonas} and {MeerKAT Team}(2016)}]{jonas2016meerkat}
\bibinfo{author}{{Jonas}, J.}, \bibinfo{author}{{MeerKAT Team}},
  \bibinfo{year}{2016}.
\newblock \bibinfo{title}{{The MeerKAT Radio Telescope}}, in:
  \bibinfo{booktitle}{MeerKAT Science: On the Pathway to the SKA},
  p.~\bibinfo{pages}{1}.
\newblock \DOIprefix\doi{10.22323/1.277.0001}.
%Type = Article
\bibitem[{{Junklewitz} et~al.(2013){Junklewitz}, {Bell}, {Selig} and
  {En{\ss}lin}}]{resolve1}
\bibinfo{author}{{Junklewitz}, H.}, \bibinfo{author}{{Bell}, M.R.},
  \bibinfo{author}{{Selig}, M.}, \bibinfo{author}{{En{\ss}lin}, T.A.},
  \bibinfo{year}{2013}.
\newblock \bibinfo{title}{{RESOLVE: A new algorithm for aperture synthesis
  imaging of extended emission in radio astronomy}}.
\newblock \bibinfo{journal}{ArXiv e-prints}
  \href{http://arxiv.org/abs/1311.5282}{{\tt arXiv:1311.5282}}.
%Type = Article
\bibitem[{Kabanikhin and Shishlenin(2019)}]{illposed2019}
\bibinfo{author}{Kabanikhin, S.}, \bibinfo{author}{Shishlenin, M.},
  \bibinfo{year}{2019}.
\newblock \bibinfo{title}{Theory and numerical methods for solving inverse and
  ill-posed problems}.
\newblock \bibinfo{journal}{Journal of Inverse and Ill-posed Problems}
  \bibinfo{volume}{27}, \bibinfo{pages}{453--456}.
\newblock \DOIprefix\doi{doi:10.1515/jiip-2019-5001}.
%Type = Article
\bibitem[{Kenyon et~al.(2024)Kenyon, Perkins, Bester, Smirnov, Russeeawon and
  Hugo}]{africanus2}
\bibinfo{author}{Kenyon, J.S.}, \bibinfo{author}{Perkins, S.J.},
  \bibinfo{author}{Bester, H.L.}, \bibinfo{author}{Smirnov, O.M.},
  \bibinfo{author}{Russeeawon, C.}, \bibinfo{author}{Hugo, B.V.},
  \bibinfo{year}{2024}.
\newblock \bibinfo{title}{{Africanus II. QuartiCal: calibrating radio
  interferometer data at scale using Numba and Dask}}.
\newblock \bibinfo{journal}{Astronomy and Computing}
  \bibinfo{volume}{submitted}.
\newblock \href{http://arxiv.org/abs/2412.10072}{{\tt arXiv:2412.10072}}.
%Type = Inproceedings
\bibitem[{Lam et~al.(2015)Lam, Pitrou and Seibert}]{lam2015}
\bibinfo{author}{Lam, S.K.}, \bibinfo{author}{Pitrou, A.},
  \bibinfo{author}{Seibert, S.}, \bibinfo{year}{2015}.
\newblock \bibinfo{title}{Numba: A llvm-based python jit compiler}, in:
  \bibinfo{booktitle}{Proceedings of the Second Workshop on the LLVM Compiler
  Infrastructure in HPC}, pp. \bibinfo{pages}{1--6}.
%Type = Article
\bibitem[{Lee et~al.(2019)Lee, Gommers, Waselewski, Wohlfahrt and
  O'Leary}]{Lee2019}
\bibinfo{author}{Lee, G.R.}, \bibinfo{author}{Gommers, R.},
  \bibinfo{author}{Waselewski, F.}, \bibinfo{author}{Wohlfahrt, K.},
  \bibinfo{author}{O'Leary, A.}, \bibinfo{year}{2019}.
\newblock \bibinfo{title}{Pywavelets: A python package for wavelet analysis}.
\newblock \bibinfo{journal}{Astronomy and Computing} \bibinfo{volume}{4},
  \bibinfo{pages}{1237}.
\newblock \DOIprefix\doi{10.21105/joss.01237}.
%Type = Misc
\bibitem[{Liaudat et~al.(2024)Liaudat, Mars, Price, Pereyra, Betcke and
  McEwen}]{liaudat2024}
\bibinfo{author}{Liaudat, T.I.}, \bibinfo{author}{Mars, M.},
  \bibinfo{author}{Price, M.A.}, \bibinfo{author}{Pereyra, M.},
  \bibinfo{author}{Betcke, M.M.}, \bibinfo{author}{McEwen, J.D.},
  \bibinfo{year}{2024}.
\newblock \bibinfo{title}{Scalable bayesian uncertainty quantification with
  data-driven priors for radio interferometric imaging}.
\newblock \href{http://arxiv.org/abs/2312.00125}{{\tt arXiv:2312.00125}}.
%Type = Book
\bibitem[{Nocedal and Wright(2006)}]{nocedal06}
\bibinfo{author}{Nocedal, J.}, \bibinfo{author}{Wright, S.J.},
  \bibinfo{year}{2006}.
\newblock \bibinfo{title}{Numerical Optimization}.
\newblock \bibinfo{edition}{second} ed., \bibinfo{publisher}{Springer},
  \bibinfo{address}{New York, NY, USA}.
%Type = Article
\bibitem[{Offringa et~al.(2014)Offringa, McKinley, Hurley-Walker, Briggs,
  Wayth, Kaplan, Bell, Feng, Neben, Hughes, Rhee, Murphy, Bhat, Bernardi,
  Bowman, Cappallo, Corey, Deshpande, Emrich, Ewall-Wice, Gaensler, Goeke,
  Greenhill, Hazelton, Hindson, Johnston-Hollitt, Jacobs, Kasper, Kratzenberg,
  Lenc, Lonsdale, Lynch, McWhirter, Mitchell, Morales, Morgan, Kudryavtseva,
  Oberoi, Ord, Pindor, Procopio, Prabu, Riding, Roshi, Shankar, Srivani,
  Subrahmanyan, Tingay, Waterson, Webster, Whitney, Williams and
  Williams}]{Offringa_2014}
\bibinfo{author}{Offringa, A.R.}, \bibinfo{author}{McKinley, B.},
  \bibinfo{author}{Hurley-Walker, N.}, \bibinfo{author}{Briggs, F.H.},
  \bibinfo{author}{Wayth, R.B.}, \bibinfo{author}{Kaplan, D.L.},
  \bibinfo{author}{Bell, M.E.}, \bibinfo{author}{Feng, L.},
  \bibinfo{author}{Neben, A.R.}, \bibinfo{author}{Hughes, J.D.},
  \bibinfo{author}{Rhee, J.}, \bibinfo{author}{Murphy, T.},
  \bibinfo{author}{Bhat, N.D.R.}, \bibinfo{author}{Bernardi, G.},
  \bibinfo{author}{Bowman, J.D.}, \bibinfo{author}{Cappallo, R.J.},
  \bibinfo{author}{Corey, B.E.}, \bibinfo{author}{Deshpande, A.A.},
  \bibinfo{author}{Emrich, D.}, \bibinfo{author}{Ewall-Wice, A.},
  \bibinfo{author}{Gaensler, B.M.}, \bibinfo{author}{Goeke, R.},
  \bibinfo{author}{Greenhill, L.J.}, \bibinfo{author}{Hazelton, B.J.},
  \bibinfo{author}{Hindson, L.}, \bibinfo{author}{Johnston-Hollitt, M.},
  \bibinfo{author}{Jacobs, D.C.}, \bibinfo{author}{Kasper, J.C.},
  \bibinfo{author}{Kratzenberg, E.}, \bibinfo{author}{Lenc, E.},
  \bibinfo{author}{Lonsdale, C.J.}, \bibinfo{author}{Lynch, M.J.},
  \bibinfo{author}{McWhirter, S.R.}, \bibinfo{author}{Mitchell, D.A.},
  \bibinfo{author}{Morales, M.F.}, \bibinfo{author}{Morgan, E.},
  \bibinfo{author}{Kudryavtseva, N.}, \bibinfo{author}{Oberoi, D.},
  \bibinfo{author}{Ord, S.M.}, \bibinfo{author}{Pindor, B.},
  \bibinfo{author}{Procopio, P.}, \bibinfo{author}{Prabu, T.},
  \bibinfo{author}{Riding, J.}, \bibinfo{author}{Roshi, D.A.},
  \bibinfo{author}{Shankar, N.U.}, \bibinfo{author}{Srivani, K.S.},
  \bibinfo{author}{Subrahmanyan, R.}, \bibinfo{author}{Tingay, S.J.},
  \bibinfo{author}{Waterson, M.}, \bibinfo{author}{Webster, R.L.},
  \bibinfo{author}{Whitney, A.R.}, \bibinfo{author}{Williams, A.},
  \bibinfo{author}{Williams, C.L.}, \bibinfo{year}{2014}.
\newblock \bibinfo{title}{wsclean: an implementation of a fast, generic
  wide-field imager for radio astronomy}.
\newblock \bibinfo{journal}{Monthly Notices of the Royal Astronomical Society}
  \bibinfo{volume}{444}, \bibinfo{pages}{606--619}.
\newblock \DOIprefix\doi{10.1093/mnras/stu1368}.
%Type = Article
\bibitem[{{Offringa} and {Smirnov}(2017)}]{offringa2017multiscale}
\bibinfo{author}{{Offringa}, A.R.}, \bibinfo{author}{{Smirnov}, O.},
  \bibinfo{year}{2017}.
\newblock \bibinfo{title}{{An optimized algorithm for multiscale wideband
  deconvolution of radio astronomical images}}.
\newblock \bibinfo{journal}{MNRAS} \bibinfo{volume}{471},
  \bibinfo{pages}{301--316}.
\newblock \DOIprefix\doi{10.1093/mnras/stx1547},
  \href{http://arxiv.org/abs/1706.06786}{{\tt arXiv:1706.06786}}.
%Type = Article
\bibitem[{Perkins et~al.(2024)Perkins, Kenyon, Andati, Bester, Smirnov and
  Hugo}]{africanus1}
\bibinfo{author}{Perkins, S.J.}, \bibinfo{author}{Kenyon, J.S.},
  \bibinfo{author}{Andati, L.A.L.}, \bibinfo{author}{Bester, H.L.},
  \bibinfo{author}{Smirnov, O.M.}, \bibinfo{author}{Hugo, B.V.},
  \bibinfo{year}{2024}.
\newblock \bibinfo{title}{{Africanus I. Scalable, distributed and efficient
  radio data processing with Dask-MS and Codex Africanus}}.
\newblock \bibinfo{journal}{Astronomy and Computing}
  \bibinfo{volume}{submitted}.
\newblock \href{http://arxiv.org/abs/2412.12052}{{\tt arXiv:2412.12052}}.
%Type = Article
\bibitem[{{Rau} and {Cornwell}(2011)}]{clean2011}
\bibinfo{author}{{Rau}, U.}, \bibinfo{author}{{Cornwell}, T.J.},
  \bibinfo{year}{2011}.
\newblock \bibinfo{title}{{A multi-scale multi-frequency deconvolution
  algorithm for synthesis imaging in radio interferometry}}.
\newblock \bibinfo{journal}{A\& A} \bibinfo{volume}{532}, \bibinfo{pages}{A71}.
\newblock \DOIprefix\doi{10.1051/0004-6361/201117104},
  \href{http://arxiv.org/abs/1106.2745}{{\tt arXiv:1106.2745}}.
%Type = Article
\bibitem[{Repetti et~al.(2017)Repetti, Birdi, Dabbech and
  Wiaux}]{repetti2017non}
\bibinfo{author}{Repetti, A.}, \bibinfo{author}{Birdi, J.},
  \bibinfo{author}{Dabbech, A.}, \bibinfo{author}{Wiaux, Y.},
  \bibinfo{year}{2017}.
\newblock \bibinfo{title}{Non-convex optimization for self-calibration of
  direction-dependent effects in radio interferometric imaging}.
\newblock \bibinfo{journal}{Monthly Notices of the Royal Astronomical Society}
  \bibinfo{volume}{470}, \bibinfo{pages}{3981--4006}.
%Type = Article
\bibitem[{Repetti and Wiaux(2021)}]{pfb3}
\bibinfo{author}{Repetti, A.}, \bibinfo{author}{Wiaux, Y.},
  \bibinfo{year}{2021}.
\newblock \bibinfo{title}{Variable metric forward-backward algorithm for
  composite minimization problems}.
\newblock \bibinfo{journal}{SIAM Journal on Optimization} \bibinfo{volume}{31},
  \bibinfo{pages}{1215--1241}.
\newblock \DOIprefix\doi{10.1137/19M1277552},
  \href{http://arxiv.org/abs/https://doi.org/10.1137/19M1277552}{{\tt
  arXiv:https://doi.org/10.1137/19M1277552}}.
%Type = Inproceedings
\bibitem[{Rocklin(2015)}]{rocklin2015}
\bibinfo{author}{Rocklin, M.}, \bibinfo{year}{2015}.
\newblock \bibinfo{title}{Dask: Parallel computation with blocked algorithms
  and task scheduling}, in: \bibinfo{editor}{Huff, K.},
  \bibinfo{editor}{Bergstra, J.} (Eds.), \bibinfo{booktitle}{Proceedings of the
  14th Python in Science Conference}, pp. \bibinfo{pages}{130 -- 136}.
%Type = Article
\bibitem[{{Roth} et~al.(2024){Roth}, {Frank}, {Bester}, {Smirnov}, {Westermann}
  and {En{\ss}lin}}]{roth2024fastresolvefastbayesianradio}
\bibinfo{author}{{Roth}, J.}, \bibinfo{author}{{Frank}, P.},
  \bibinfo{author}{{Bester}, H.L.}, \bibinfo{author}{{Smirnov}, O.M.},
  \bibinfo{author}{{Westermann}, R.}, \bibinfo{author}{{En{\ss}lin}, T.A.},
  \bibinfo{year}{2024}.
\newblock \bibinfo{title}{{fast-resolve: Fast Bayesian radio interferometric
  imaging}}.
\newblock \bibinfo{journal}{A\& A} \bibinfo{volume}{690},
  \bibinfo{pages}{A387}.
\newblock \DOIprefix\doi{10.1051/0004-6361/202451107},
  \href{http://arxiv.org/abs/2406.09144}{{\tt arXiv:2406.09144}}.
%Type = Article
\bibitem[{{Sault} et~al.(1996){Sault}, {Hamaker} and
  {Bregman}}]{hamaker1996understanding2}
\bibinfo{author}{{Sault}, R.J.}, \bibinfo{author}{{Hamaker}, J.P.},
  \bibinfo{author}{{Bregman}, J.D.}, \bibinfo{year}{1996}.
\newblock \bibinfo{title}{{Understanding radio polarimetry. II. Instrumental
  calibration of an interferometer array.}}
\newblock \bibinfo{journal}{A\& As} \bibinfo{volume}{117},
  \bibinfo{pages}{149--159}.
%Type = Inproceedings
\bibitem[{Schilizzi et~al.(2008)Schilizzi, Dewdney and
  Lazio}]{schilizzi2008square}
\bibinfo{author}{Schilizzi, R.T.}, \bibinfo{author}{Dewdney, P.E.},
  \bibinfo{author}{Lazio, T.J.W.}, \bibinfo{year}{2008}.
\newblock \bibinfo{title}{The square kilometre array}, in:
  \bibinfo{booktitle}{SPIE Astronomical Telescopes+ Instrumentation},
  \bibinfo{organization}{International Society for Optics and Photonics}. pp.
  \bibinfo{pages}{70121I--70121I}.
%Type = Article
\bibitem[{{Smirnov}(2011a)}]{smirnov2011revisiting1}
\bibinfo{author}{{Smirnov}, O.M.}, \bibinfo{year}{2011}a.
\newblock \bibinfo{title}{{Revisiting the radio interferometer measurement
  equation. I. A full-sky Jones formalism}}.
\newblock \bibinfo{journal}{A\& A} \bibinfo{volume}{527},
  \bibinfo{pages}{A106}.
\newblock \DOIprefix\doi{10.1051/0004-6361/201016082},
  \href{http://arxiv.org/abs/1101.1764}{{\tt arXiv:1101.1764}}.
%Type = Article
\bibitem[{{Smirnov}(2011b)}]{smirnov2011revisiting2}
\bibinfo{author}{{Smirnov}, O.M.}, \bibinfo{year}{2011}b.
\newblock \bibinfo{title}{{Revisiting the radio interferometer measurement
  equation. II. Calibration and direction-dependent effects}}.
\newblock \bibinfo{journal}{A\& A} \bibinfo{volume}{527},
  \bibinfo{pages}{A107}.
\newblock \DOIprefix\doi{10.1051/0004-6361/201116434},
  \href{http://arxiv.org/abs/1101.1765}{{\tt arXiv:1101.1765}}.
%Type = Article
\bibitem[{Smirnov et~al.(2024)Smirnov, Makhathini, Kenyon, Bester, Perkins,
  Ramaila and Hugo}]{africanus4}
\bibinfo{author}{Smirnov, O.M.}, \bibinfo{author}{Makhathini, S.},
  \bibinfo{author}{Kenyon, J.S.}, \bibinfo{author}{Bester, H.L.},
  \bibinfo{author}{Perkins, S.J.}, \bibinfo{author}{Ramaila, A.J.T.},
  \bibinfo{author}{Hugo, B.V.}, \bibinfo{year}{2024}.
\newblock \bibinfo{title}{{Africanus IV. The Stimela2 framework: scalable and
  reproducible workflows, from local to cloud compute}}.
\newblock \bibinfo{journal}{Astronomy and Computing}
  \bibinfo{volume}{submitted}.
\newblock \href{http://arxiv.org/abs/2412.10080}{{\tt arXiv:2412.10080}}.
%Type = Article
\bibitem[{{Smirnov} et~al.(2024){Smirnov}, {Stappers}, {Tasse}, {Bester},
  {Bignall}, {Walker}, {Caleb}, {Rajwade}, {Buchner}, {Woudt}, {Ivchenko},
  {Roth}, {Noordam} and {Camilo}}]{smirnov2024}
\bibinfo{author}{{Smirnov}, O.M.}, \bibinfo{author}{{Stappers}, B.W.},
  \bibinfo{author}{{Tasse}, C.}, \bibinfo{author}{{Bester}, H.L.},
  \bibinfo{author}{{Bignall}, H.}, \bibinfo{author}{{Walker}, M.A.},
  \bibinfo{author}{{Caleb}, M.}, \bibinfo{author}{{Rajwade}, K.M.},
  \bibinfo{author}{{Buchner}, S.}, \bibinfo{author}{{Woudt}, P.},
  \bibinfo{author}{{Ivchenko}, M.}, \bibinfo{author}{{Roth}, L.},
  \bibinfo{author}{{Noordam}, J.E.}, \bibinfo{author}{{Camilo}, F.},
  \bibinfo{year}{2024}.
\newblock \bibinfo{title}{{The RATT PARROT: serendipitous discovery of a
  peculiarly scintillating pulsar in MeerKAT imaging observations of the Great
  Saturn - Jupiter Conjunction of 2020. I. Dynamic imaging and data analysis}}.
\newblock \bibinfo{journal}{MNRAS} \bibinfo{volume}{528},
  \bibinfo{pages}{6517--6537}.
\newblock \DOIprefix\doi{10.1093/mnras/stae303}.
%Type = Article
\bibitem[{Sob et~al.(2021)Sob, Bester, Smirnov, Kenyon and
  Russeeawon}]{sob2021solints}
\bibinfo{author}{Sob, U.M.}, \bibinfo{author}{Bester, H.L.},
  \bibinfo{author}{Smirnov, O.M.}, \bibinfo{author}{Kenyon, J.S.},
  \bibinfo{author}{Russeeawon, C.}, \bibinfo{year}{2021}.
\newblock \bibinfo{title}{{Solution intervals considered harmful: on the
  optimality of radio interferometric gain solutions}}.
\newblock \bibinfo{journal}{Monthly Notices of the Royal Astronomical Society}
  \bibinfo{volume}{504}, \bibinfo{pages}{1714--1732}.
\newblock \DOIprefix\doi{10.1093/mnras/stab928}.
%Type = Article
\bibitem[{{Sutter} et~al.(2014){Sutter}, {Wandelt}, {McEwen}, {Bunn},
  {Karakci}, {Korotkov}, {Timbie}, {Tucker} and {Zhang}}]{sutter2014}
\bibinfo{author}{{Sutter}, P.M.}, \bibinfo{author}{{Wandelt}, B.D.},
  \bibinfo{author}{{McEwen}, J.D.}, \bibinfo{author}{{Bunn}, E.F.},
  \bibinfo{author}{{Karakci}, A.}, \bibinfo{author}{{Korotkov}, A.},
  \bibinfo{author}{{Timbie}, P.}, \bibinfo{author}{{Tucker}, G.S.},
  \bibinfo{author}{{Zhang}, L.}, \bibinfo{year}{2014}.
\newblock \bibinfo{title}{{Probabilistic image reconstruction for radio
  interferometers}}.
\newblock \bibinfo{journal}{MNRAS} \bibinfo{volume}{438},
  \bibinfo{pages}{768--778}.
\newblock \DOIprefix\doi{10.1093/mnras/stt2244},
  \href{http://arxiv.org/abs/1309.1469}{{\tt arXiv:1309.1469}}.
%Type = Article
\bibitem[{Terris et~al.(2022)Terris, Dabbech, Tang and Wiaux}]{terris2022}
\bibinfo{author}{Terris, M.}, \bibinfo{author}{Dabbech, A.},
  \bibinfo{author}{Tang, C.}, \bibinfo{author}{Wiaux, Y.},
  \bibinfo{year}{2022}.
\newblock \bibinfo{title}{{Image reconstruction algorithms in radio
  interferometry: From handcrafted to learned regularization denoisers}}.
\newblock \bibinfo{journal}{Monthly Notices of the Royal Astronomical Society}
  \bibinfo{volume}{518}, \bibinfo{pages}{604--622}.
\newblock \DOIprefix\doi{10.1093/mnras/stac2672}.
%Type = Article
\bibitem[{{THE CASA TEAM} et~al.(2022){THE CASA TEAM}, {Bean}, {Bhatnagar},
  {Castro}, {Donovan Meyer}, {Emonts}, {Garcia}, {Garwood}, {Golap}, {Gonzalez
  Villalba}, {Harris}, {Hayashi}, {Hoskins}, {Hsieh}, {Jagannathan},
  {Kawasaki}, {Keimpema}, {Kettenis}, {Lopez}, {Marvil}, {Masters},
  {McNichols}, {Mehringer}, {Miel}, {Moellenbrock}, {Montesino}, {Nakazato},
  {Ott}, {Petry}, {Pokorny}, {Raba}, {Rau}, {Schiebel}, {Schweighart},
  {Sekhar}, {Shimada}, {Small}, {Steeb}, {Sugimoto}, {Suoranta}, {Tsutsumi},
  {van Bemmel}, {Verkouter}, {Wells}, {Xiong}, {Szomoru}, {Griffith},
  {Glendenning} and {Kern}}]{casa2022}
\bibinfo{author}{{THE CASA TEAM}}, \bibinfo{author}{{Bean}, B.},
  \bibinfo{author}{{Bhatnagar}, S.}, \bibinfo{author}{{Castro}, S.},
  \bibinfo{author}{{Donovan Meyer}, J.}, \bibinfo{author}{{Emonts}, B.},
  \bibinfo{author}{{Garcia}, E.}, \bibinfo{author}{{Garwood}, R.},
  \bibinfo{author}{{Golap}, K.}, \bibinfo{author}{{Gonzalez Villalba}, J.},
  \bibinfo{author}{{Harris}, P.}, \bibinfo{author}{{Hayashi}, Y.},
  \bibinfo{author}{{Hoskins}, J.}, \bibinfo{author}{{Hsieh}, M.},
  \bibinfo{author}{{Jagannathan}, P.}, \bibinfo{author}{{Kawasaki}, W.},
  \bibinfo{author}{{Keimpema}, A.}, \bibinfo{author}{{Kettenis}, M.},
  \bibinfo{author}{{Lopez}, J.}, \bibinfo{author}{{Marvil}, J.},
  \bibinfo{author}{{Masters}, J.}, \bibinfo{author}{{McNichols}, A.},
  \bibinfo{author}{{Mehringer}, D.}, \bibinfo{author}{{Miel}, R.},
  \bibinfo{author}{{Moellenbrock}, G.}, \bibinfo{author}{{Montesino}, F.},
  \bibinfo{author}{{Nakazato}, T.}, \bibinfo{author}{{Ott}, J.},
  \bibinfo{author}{{Petry}, D.}, \bibinfo{author}{{Pokorny}, M.},
  \bibinfo{author}{{Raba}, R.}, \bibinfo{author}{{Rau}, U.},
  \bibinfo{author}{{Schiebel}, D.}, \bibinfo{author}{{Schweighart}, N.},
  \bibinfo{author}{{Sekhar}, S.}, \bibinfo{author}{{Shimada}, K.},
  \bibinfo{author}{{Small}, D.}, \bibinfo{author}{{Steeb}, J.W.},
  \bibinfo{author}{{Sugimoto}, K.}, \bibinfo{author}{{Suoranta}, V.},
  \bibinfo{author}{{Tsutsumi}, T.}, \bibinfo{author}{{van Bemmel}, I.M.},
  \bibinfo{author}{{Verkouter}, M.}, \bibinfo{author}{{Wells}, A.},
  \bibinfo{author}{{Xiong}, W.}, \bibinfo{author}{{Szomoru}, A.},
  \bibinfo{author}{{Griffith}, M.}, \bibinfo{author}{{Glendenning}, B.},
  \bibinfo{author}{{Kern}, J.}, \bibinfo{year}{2022}.
\newblock \bibinfo{title}{{CASA, the Common Astronomy Software Applications for
  Radio Astronomy}}.
\newblock \bibinfo{journal}{arXiv e-prints} ,
  \bibinfo{pages}{arXiv:2210.02276}\href{http://arxiv.org/abs/2210.02276}{{\tt
  arXiv:2210.02276}}.
%Type = Article
\bibitem[{{Thouvenin} et~al.(2023){Thouvenin}, {Dabbech}, {Jiang}, {Abdulaziz},
  {Thiran}, {Jackson} and {Wiaux}}]{thouvenin2023}
\bibinfo{author}{{Thouvenin}, P.A.}, \bibinfo{author}{{Dabbech}, A.},
  \bibinfo{author}{{Jiang}, M.}, \bibinfo{author}{{Abdulaziz}, A.},
  \bibinfo{author}{{Thiran}, J.P.}, \bibinfo{author}{{Jackson}, A.},
  \bibinfo{author}{{Wiaux}, Y.}, \bibinfo{year}{2023}.
\newblock \bibinfo{title}{{Parallel faceted imaging in radio interferometry via
  proximal splitting (Faceted HyperSARA) - II. Code and real data proof of
  concept}}.
\newblock \bibinfo{journal}{MNRAS} \bibinfo{volume}{521},
  \bibinfo{pages}{20--34}.
\newblock \DOIprefix\doi{10.1093/mnras/stac3175},
  \href{http://arxiv.org/abs/2209.07604}{{\tt arXiv:2209.07604}}.
%Type = Article
\bibitem[{{van Diepen}(2015)}]{vandiepen2015}
\bibinfo{author}{{van Diepen}, G.N.J.}, \bibinfo{year}{2015}.
\newblock \bibinfo{title}{{Casacore Table Data System and its use in the
  MeasurementSet}}.
\newblock \bibinfo{journal}{Astronomy and Computing} \bibinfo{volume}{12},
  \bibinfo{pages}{174--180}.
\newblock \DOIprefix\doi{10.1016/j.ascom.2015.06.002}.
%Type = Article
\bibitem[{{van Haarlem} et~al.(2013)}]{vanhaarlem2013}
\bibinfo{author}{{van Haarlem}, M.P.}, et~al., \bibinfo{year}{2013}.
\newblock \bibinfo{title}{{LOFAR: The LOw-Frequency ARray}}.
\newblock \bibinfo{journal}{A\&A} \bibinfo{volume}{556}, \bibinfo{pages}{A2}.
\newblock \DOIprefix\doi{10.1051/0004-6361/201220873},
  \href{http://arxiv.org/abs/1305.3550}{{\tt arXiv:1305.3550}}.
%Type = Book
\bibitem[{Van~Rossum and Drake(2009)}]{vanrossum2009}
\bibinfo{author}{Van~Rossum, G.}, \bibinfo{author}{Drake, F.L.},
  \bibinfo{year}{2009}.
\newblock \bibinfo{title}{Python 3 Reference Manual}.
\newblock \bibinfo{publisher}{CreateSpace}, \bibinfo{address}{Scotts Valley,
  CA}.
%Type = Article
\bibitem[{Virtanen et~al.(2020)}]{Virtanen2020}
\bibinfo{author}{Virtanen, P.}, et~al., \bibinfo{year}{2020}.
\newblock \bibinfo{title}{{{SciPy} 1.0: Fundamental Algorithms for Scientific
  Computing in Python}}.
\newblock \bibinfo{journal}{Nature Methods} \bibinfo{volume}{17},
  \bibinfo{pages}{261--272}.
\newblock \DOIprefix\doi{10.1038/s41592-019-0686-2}.
%Type = Article
\bibitem[{{Wells} et~al.(1981){Wells}, {Greisen} and {Harten}}]{Wells1981}
\bibinfo{author}{{Wells}, D.C.}, \bibinfo{author}{{Greisen}, E.W.},
  \bibinfo{author}{{Harten}, R.H.}, \bibinfo{year}{1981}.
\newblock \bibinfo{title}{{FITS - a Flexible Image Transport System}}.
\newblock \bibinfo{journal}{A\&A} \bibinfo{volume}{44}, \bibinfo{pages}{363}.
%Type = Article
\bibitem[{{Wilber} et~al.(2023){Wilber}, {Dabbech}, {Jackson} and
  {Wiaux}}]{wilber2023}
\bibinfo{author}{{Wilber}, A.G.}, \bibinfo{author}{{Dabbech}, A.},
  \bibinfo{author}{{Jackson}, A.}, \bibinfo{author}{{Wiaux}, Y.},
  \bibinfo{year}{2023}.
\newblock \bibinfo{title}{{Scalable precision wide-field imaging in radio
  interferometry: I. uSARA validated on ASKAP data}}.
\newblock \bibinfo{journal}{MNRAS} \bibinfo{volume}{522},
  \bibinfo{pages}{5558--5575}.
\newblock \DOIprefix\doi{10.1093/mnras/stad1351},
  \href{http://arxiv.org/abs/2302.14148}{{\tt arXiv:2302.14148}}.
%Type = Article
\bibitem[{Ye et~al.(2021)Ye, Gull, Tan and Nikolic}]{Ye_2021}
\bibinfo{author}{Ye, H.}, \bibinfo{author}{Gull, S.F.}, \bibinfo{author}{Tan,
  S.M.}, \bibinfo{author}{Nikolic, B.}, \bibinfo{year}{2021}.
\newblock \bibinfo{title}{High accuracy wide-field imaging method in radio
  interferometry}.
\newblock \bibinfo{journal}{Monthly Notices of the Royal Astronomical Society}
  \bibinfo{volume}{510}, \bibinfo{pages}{4110–4125}.
\newblock \DOIprefix\doi{10.1093/mnras/stab3548}.

\end{thebibliography}

\appendix

\section{Calibration}
\label{sec:calibration}
The first step of a typical data processing pipeline for radio interferometry is to calibrate the instrument against a known sky. This makes it possible to get a handle on the main systematics affecting the observation. Typically, at least for total intensity imaging at L-band, it is sufficient to parametrise the diagonal elements of the Jones terms affecting the observation as follows:
\begin{equation}
    J_p(t, \nu) = G_p(t) K_p(t, \nu) B_p(\nu),
    \label{jchain}
\end{equation}
where the complex valued term $B_p(\nu)$ is often referred to as the band-pass as it models the frequency response of the antenna, $G_p(t)$ is a complex time dependent gain that captures residual time variability in the electronic receiver system and $K_p(t, \nu) = \exp(-\imath d(t) (\nu -\nu_c))$, with $\nu_c$ the central frequency in the band and where $d(t)$ is referred to as the delay, is a phase only term specifically accounting for the fact that there might be time-dependent variability in the path length that the signal has to traverse in going from the individual receivers to the correlator\footnote{This uses the fact that a lag in the time domain corresponds to a complex exponential in the Fourier domain i.e. $f(t-t_0) \leftrightarrow F(\nu) \exp(-2 \pi \imath t_0 \nu)$ for Fourier pairs $f$ and $F$.}. With the calibration model specified, \qcal\ can be used to solve for the calibration parameters. The exact implementation details will not concern us here, they are given in the accompanying paper \cite{africanus2}. The important thing to note is that \qcal\ splits the calibration problem into distinct time and frequency chunks that do not necessarily align with how the imaging problem is partitioned. This allows it to compute the calibration solutions in a highly efficient and distributed manner. Gain solution intervals \citep[see e.g.][]{sob2021solints}) are used to limit the number of degrees of freedom for the calibration problem and chunks sizes should ideally be chosen to align with the solution intervals. In this work, we solve for the above three terms with a time interval the length of an entire calibrator scan and use a frequency interval the width of the single channel for the band-pass\footnote{For the observation considered in $\S$~\ref{sec:results} the primary calibrator was sufficiently close to the target for the entire duration of the observation so there is only a single calibrator field.}. The frequency interval for $K$ is set to the full bandwidth of the data.

Once the gains on the calibrator field have been determined they need to be transferred to the target field. This is done by straightforward linear interpolation over the target scans (performing nearest neighbour extrapolation if required) which provides starting estimates for the gains and therefore allows self-calibration \cite{cornwell1981new} to commence. In order to make the gains available for use in the imager, they are combined into a single effective net gain, $\hat{N}_p(t,\nu)$ say, that is interpolated to the full time and frequency resolution of the data. After a round of imaging, performed as explained in the next section, the gain estimates can be refined using the resulting model. Note that the model can be interpolated to an arbitrary frequency resolution, enabling better calibration accuracy than can be achieved with the coarsely discretised model resulting from discretising the operator as shown in \eqref{immeasop}.

At this stage of the reduction, given that the model is probably not yet complete, conventional wisdom dictates that we avoid solving for amplitude effects as these can too easily absorb unmodelled flux. Also, as is typically the case at L-band, the main effects that still need to be calibrated for are residual delay and time dependent phase drifts. Hence the calibration model that is used after the first round of imaging is given by
\begin{equation}
    J_p(t, \nu) = \hat{N}_p(t, \nu) D_p(t, \nu),
    \label{jchain2}
\end{equation}
where $\hat{N}_p(t,\nu)$ is the fixed initial net gain obtained from transferring the calibrator solutions onto the target field and
\begin{equation}
    D_p(t, \nu) = \exp\left(\imath (d(t) (\nu - \nu_c) +  \theta(t)) \right),
    \label{delpoff}
\end{equation}
captures the sought after refinement in delay and phase. We used time intervals the length of a single integration time and again set the frequency interval to the full bandwidth of the data. The gains are again combined into a single effective net gain term that is used during the next round of imaging.

After the second round of imaging the model is complete enough to safely solve for both amplitude and phase effects. Thus we use a final calibration model of the form
\begin{equation}
    J_p(t, \nu) = \hat{N}_p(t, \nu) C_p(t, \nu) D_p(t, \nu),
    \label{jchainf}
\end{equation}
where $\hat{N}_p(t,\nu)$ and $D_p(t, \nu)$ are the same as before and $C_p(t, \nu)$ is a complex valued gain. The solution intervals for $C_p(t, \nu)$ have to be chosen carefully to achieve a good balance between fitting the residual gain errors without over-fitting the data. We found that a time interval of 75 integrations and a frequency interval of 256 channels resulted in reduced $\chi^2$ values per chunk that average to approximately one. Further refining the $D_p(t, \nu)$ solutions after solving for $C_p(t, \nu)$ gave a further improvement, decreasing the spread in the reduced $\chi^2$ values and visibly reducing the background noise in the resulting images.

The final calibration model \eqref{jchainf} was arrived at by trial and error. Some attempts were made to refine the model further but none of our experiments yielded a worthwhile improvement in the final images. It is possible that the calibration strategy becomes fundamentally limited by the use of maximum likelihood calibration in conjunction with solution intervals. This may explain some of the limitations of the results presented in $\S$~\ref{sec:results}. Further improvements to the calibration strategy might be possible by incorporating certain physical constraint (e.g. smoothness of the band-pass) and by solving the imaging and calibration problems in a more statistically consistent way \citep[see e.g.][]{repetti2017non}). This will be investigated in future research.

\onecolumn
\section{\stimela\ recipes}
This section provides the commands and \stimela\ recipes used to generate the results presented in this paper. All of the \texttt{cab} definitions (see $\S$~4 of \cite{africanus4}) required by the recipes are provided in the \textsc{cult-cargo}\footnote{\url{https://github.com/caracal-pipeline/cult-cargo}} package. We utilised version 0.1.3 for all experiments.

\subsection{\wsclean:}
\label{sec:wscommand}
The command used to generate the \wsclean\ results shown in Figures~\ref{fig:pfbvwsc}-\ref{fig:eso007} is
\begin{lstlisting}[style=bash]
stimela run wscimage.yaml image obs=eso nthreads=64 weight=\[briggs,-0.3\]
\end{lstlisting}
This command invokes the following \stimela\ recipe:
\yamlfromfile{wscimage.yaml}
The corresponding \wsclean\ command is
\begin{lstlisting}[style=bash]
wsclean -j 64 -v -log-time -weight briggs -0.3 -name output/wsclean/run1 -size 7560 7560 -scale 0.953795asec -channels-out 20 -no-update-model-required -use-wgridder -niter 1000000 -nmiter 10 -auto-threshold 1.0 -auto-mask 3.5 -mgain 0.85 -join-channels -multiscale -multiscale-max-scales 6 -multiscale-gain 0.15 -fit-spectral-pol 4 -data-column DATA msdir/ms1_target_ave4chan_bdafov2d98.ms msdir/ms2_target_ave4chan_bdafov2d98.ms
\end{lstlisting}
The command used to generate the \wsclean\ results shown in Figure~\ref{fig:pfbvwsc_sgra} is
\begin{lstlisting}[style=bash]
stimela run wscimage.yaml image obs=sgra nthreads=64 weight=\[briggs,-1.0\]
\end{lstlisting}
This invokes the same \stimela\ recipe which has some of the observation specific parameters hardcoded into it.

\subsection{\pfb:}
\label{sec:recipes}
The command used to generate the \pfb\ results shown in Figures~\ref{fig:pfbvwsc}-\ref{fig:eso007} is
\begin{lstlisting}[style=bash]
stimela run pfbimage.yaml image obs=esofull basedir=outputs
\end{lstlisting}
This command invokes the following \stimela\ recipe
\yamlfromfile{pfbimage.yaml}
The command used to generate the \pfb\ results shown in Figure~\ref{fig:pfbf} is
\begin{lstlisting}[style=bash]
    stimela run pfbimage.yaml image obs=esohi basedir=outputs
\end{lstlisting}
The command used to run \pfb\ on \aws\ is
\begin{lstlisting}[style=bash]
stimela run pfbimage.yaml kubeconfig.yaml image obs=esohi basedir=s3://rarg-test-binface/ESO137 log-directory=/mnt/data/pfb-test/logs
\end{lstlisting}
This command invokes the same stimela recipe augmented with the {\sf kubeconfig.yaml} file (provided in Appendix B of \cite{africanus4}) which defines the required \kubernetes\ configuration to deploy the workflow on \aws{}. This file also defines the cabs and steps required to retrieve the component model from S3 and compare it with the one produced locally.

\end{document}